\documentclass[11pt,a4paper]{article}
\pdfoutput=1
\usepackage{jheppub}
\usepackage[FIGTOPCAP]{subfigure}
\usepackage{amsmath}
\usepackage{amsfonts}
\usepackage{amssymb}
\usepackage{tensor}
\usepackage{bbm}
\usepackage{multirow}
\usepackage{color}

\title{Drag phenomena from holographic massive gravity}
\author[a]{Matteo Baggioli,\note{E-mail address: mbaggioli@ifae.es}}
\author[b,c]{Daniel K.~Brattan.\note{E-mail address: danny.brattan@gmail.com}}
\affiliation[a]{Departament de F\'{i}sica and IFAE, Universitat Aut\`{o}noma de Barcelona, \\ Bellaterra, 08193, Barcelona, Spain.}
\affiliation[b]{Physics Department, Technion- Israel Institute of Technology,
Technion City - Haifa, \\ 32000, Israel.}
\affiliation[c]{Department of Mathematics-Physics-Computer Science, University
of Haifa at Oranim, \\ Qiryat Tivon, 36006, Israel.}

\abstract{We consider the motion of point particles in a strongly coupled field theory with broken translation invariance.  We obtain the energy and momentum loss rates and drag coefficients for a class of such particles by solving for the motion of classical strings in holographic massive gravity. At low temperatures compared to the graviton mass the behaviour of the string is controlled by the appearance of an exotic ground state with non-zero entropy at zero temperature. Additionally we find an upper bound on the diffusion constant for a collection of these particles which is saturated when the mass of the graviton goes to zero.}

\begin{document}

\maketitle

{\ Holography is an extremely successful tool for computing physical observables in strongly coupled theories by considering weak perturbations of a gravity theory in one higher dimension \cite{Maldacena:1997re,Gubser:1998bc,Witten:1998qj}. Recently, increasing interest and efforts have been devoted towards modelling condensed matter systems in an attempt to learn about real world strongly coupled systems.}

{\ To more closely approach real materials incorporating momentum dissipation mechanisms in the gravity framework is essential. Without such mechanisms weakly coupled and holographic computations often lead to unrealistic results such as the divergence of the optical conductivity at zero frequency \cite{Hartnoll:2009sz}. Several methods to introduce non-conservation of momentum have been proposed such as turning on a lattice distortion at the boundary (either through the modulation of the chemical potential or a scalar source) \cite{Horowitz:2012ky,Horowitz:2012gs,Chesler:2013qla}, the use of probe brane models \cite{Karch:2007pd,Hartnoll:2009ns}, helical lattices \cite{Donos:2014oha,Donos:2014gya}, striped phases \cite{Ling:2014laa} and finally disorder \cite{Lucas:2014zea}.}

{\ An alternative method that has proven to be a very convenient way of incorporating momentum relaxation is the introduction of a graviton mass \cite{Vegh:2013sk}. Following the original work many generalizations have been pursued with the common aim of having a complete description of strongly coupled metals, insulators and the transitions between them. For example, models with linearly sourced scalars were introduced in \cite{Andrade:2013gsa} and further developed in \cite{Taylor:2014tka,Baggioli:2014roa,Baggioli:2015zoa}. A slightly richer model than the linear scalar uses a dilaton field \cite{Gouteraux:2014hca} while others have exploited global symmetries \cite{Donos:2013eha}.}

{\ While the framework of massive gravity reproduces Drude peaks \cite{Blake:2013bqa} and the metal-insulator transitions associated with broken translation invariance, the precise interpretation of the deformation in terms of which operators we turn on in the boundary theory is unclear. For example, in many condensed matter systems non-conservation of momentum can be associated with several phenomena such as the existence of phonons or the introduction of impurities. We shall find, by examining the heat capacity and the drag coefficient, that turning on a graviton mass cannot be uniquely identified with either of these condensed matter phenomena alone. Nonetheless we do see features generic to our large class of massive gravity models, in particular, a pronounced peak in the diffusion constant.}

{\ To try and illuminate the role of graviton mass we return to one of the simplest models in condensed matter theory: the Drude model. This describes the motion of a probe particle interacting with a medium. Generally a moving probe particle loses energy and momentum to the surrounding material leading to an effective viscous drag. In a weakly coupled theory momentum is typically lost by two body collisions and the production of massless particles via bremsstrahlung. Examining the drag coefficient can generically give clues as to which mechanism is dominant, what types of collisions are occuring and the nature of the emitted massless particles.}

{\ If the medium has broken translation invariance, either by the introduction of impurities  or periodic potentials, we expect there to exist additional channels into which the probe particle can lose energy. It is important to note that the behaviour of phonons and impurities in a material are well understood in the condensed matter community. As such holographic studies investigating these phenomena are unlikely to yield new physics.} 

{\ In \cite{Gubser:2006bz,Herzog:2006gh} a probe string is used to calculate the drag on a charged particle due to thermal Yang-Mills. The string is a probe of the spacetime and experiences momentum loss to the background without significantly affecting the latter's energy. When the applied force is due to an electric field this is the holographic analogue of the Drude model. One endpoint of the string enters the black hole horizon, which describes the thermal state of the boundary field theory and acts to absorb energy and momentum supplied to the other endpoint. The second endpoint of the string in \cite{Gubser:2006bz,Herzog:2006gh} is attached to a D7-brane. One can compute the diffusion constant for quark flavour \cite{Herzog:2006gh},
  \begin{eqnarray}
   D = \frac{2}{\pi \sqrt{\lambda} T} \; 
  \end{eqnarray}
where $\lambda$ is the 't Hooft coupling, by considering the motion of the free string endpoint after it is given some initial impulse. In this paper we shall compute the diffusion constants for point particles in the field theories dual to holographic massive gravity noting that these are distinct from the $U(1)$ charge diffusion constants typically considered \cite{Vegh:2013sk,Taylor:2014tka,Baggioli:2014roa,Donos:2014oha,Donos:2014gya,Ling:2014laa,Gouteraux:2014hca,Donos:2013eha,Lucas:2014zea}.}

{\ We shall perform string drag calculations in a particular class of massive gravity models. To study the effect of momentum loss one would ideally like to study the motion of a string in the fully back-reacted spacetime dual to a boundary theory with a lattice. Doing this however requires a complex numerical calculation for the metric which obscures some of the physics. We could instead consider a model like that discussed in \cite{Blake:2013owa} which uses a weak scalar source in an Einstein-Maxwell-scalar theory to produce a modulated lattice. As was noted however the effect of this weak lattice is to give the graviton a mass precisely in the manner of \cite{Vegh:2013sk}. Since we are interested in the generic features of drag with momentum loss we shall allow arbitrary potentials for the translation breaking scalars which encode generic massive gravity theories. Thus we are taking a bottom-up approach. No consistent UV completion of these frameworks is known so far and determining such a string embedding represents an important research programme.}

{\ As the physics we are interested in concerns IR features of the theory, knowing the UV stringy completion of our model is not necessary. It is very likely that a generic string theory embedding would contain more degrees of freedom, such as a dilaton field. We would expect the dilaton, which could couple to our Stuckelberg sector, to have a radial running. Non-conformality of the boundary theory and its consequences for the drag mechanism, due to the presence of homogeneous and isotropic (in the boundary theory directions) dilaton, have already been investigated in \cite{Gursoy:2010aa}. Its effects in the context of holographic theories with momentum dissipation have also been determined \cite{Gouteraux:2014hca, Kiritsis:2015oxa, Donos:2013eha, Donos:2014uba}. In particular, a strong dilaton can greatly modify the IR solutions of the system leading to new (often insulating and gapped) phases of matter. In order to separate the consequences of breaking translational invariance from those due to the running couplings, whose effect we already know, we have assumed that the background dilaton is vanishing.}

{\ We begin the paper by discussing the theory of drag and its relevance to the Drude model of charge transport. Subsequently we give a brief description of one model of massive gravity and discuss contributions to the heat capacity at constant volume. We then compute the drag coefficient and energy/momentum loss rates to the horizon for a probe string. At low temperatures with respect to the graviton mass we find that these observables are controlled by the formation of an exotic ground state with non-zero entropy at zero temperature. Our results also indicate the existence of an upper bound on the diffusion constant of a collection of the probe particles.\\[0.1cm]}

\section{The theory of drag}

{\ Consider a particle moving with spatial momentum $\vec{p}$ in a viscous medium under a driving force $\vec{f}$. We model the motion of this particle using the drag equation
  \begin{eqnarray}
    \label{Eq:DragEquation}
    \frac{d\vec{p}}{dt} = - \gamma \vec{p} + \vec{f}
  \end{eqnarray}
where $\gamma$ is called the drag coefficient. We shall generally examine situations where the speed of the particle is low enough that its motion can be considered non-relativistic so that the dispersion relation takes the form
  \begin{eqnarray}
    E = E_{\mathrm{static}}(T) + \frac{\vec{p}^2}{2 M_{\mathrm{eff}}(T)} + \mathcal{O}^{4}(\vec{p}) \; ,
  \end{eqnarray}
where $T$ is the temperature of the medium. In this case we define the spatial momentum to be $\vec{p}=M_{\mathrm{eff}} \vec{v}$ and for steady state behaviour at constant velocity we find $\vec{v}=\vec{f}/(\gamma M_{\mathrm{eff}})$. We are particularly interested in determining the drag coefficient of this particle.}

{\ Observing the stationary velocity the particle achieves under an applied force $\vec{f}$ allows us to determine the combination $\gamma M_{\mathrm{eff}}$. Moreover if we assume the particle is charged and the applied force is due to an electric field then \eqref{Eq:DragEquation} is exactly the Drude model. The DC conductivity\footnote{In holographic theories with a $U(1)$ charge sector $\sigma_{\mathrm{DC}}$ can be shown to consist of two terms \cite{Gouteraux:2014hca}:
  \begin{equation}
    \sigma_{\mathrm{DC}}=\sigma_{\mathrm{pair}} + \sigma_{\mathrm{dissipation}} \; . 
  \end{equation}
The first term is due to quantum pair creation and can be non-zero at zero charge density. It corresponds to the conformal part of the conductivity. The second contribution is due to dissipative mechanisms and its value in the Drude model studied in undergraduate condensed matter courses is given by \eqref{Eq:SigmaDC}.} of the Drude model is well known,
  \begin{eqnarray}
    \label{Eq:SigmaDC}
    \sigma_{DC} = \frac{q^2\, n\, \tau}{M_{\mathrm{eff}}}
  \end{eqnarray}
where $\tau=\gamma^{-1}$ is called the relaxation time, $q$ is the unit of fundamental charge and $n$ is the density of charge carriers. To obtain $M_{\mathrm{eff}}$ we then consider the late time unforced behaviour of the particle with some initial momentum $\vec{p}(0)$ and solve the drag equation \eqref{Eq:DragEquation} to find
  \begin{eqnarray}
   \vec{p}(t) = \vec{p}(0) \exp(-\gamma t) \; . 
  \end{eqnarray}
A measurement of the ratio $d_{t} \vec{v}/ \vec{v}$ will then allow us to determine $\gamma$ and subsequently $M_{\mathrm{eff}}$.}

{\ Moreover, if the force in \eqref{Eq:DragEquation} contains a random element we can consider the Brownian motion of the particle through the medium. Under suitable assumptions on the nature of the statistical force it can be shown that the diffusion constant $D$ is given by $D=T/(\gamma M_{\mathrm{eff}})$ (which is called the Einstein relation). Once again if the particle has charge $q$ and we place it in an external electric field $E$ it will accelerate. At late times the velocity of that particle is $\langle v \rangle_{\mathrm{terminal}} = \frac{q E D}{T}$. Defining the mobility $\mu$ by $\langle v \rangle_{\mathrm{terminal}} = \mu (q E)$ we find
  \begin{eqnarray}
   \label{Eq:Mobilitydefinition}
   \mu = \frac{D}{T} \; ,  \qquad \mu^{-1} = \gamma M_{\mathrm{eff}} \; , 
  \end{eqnarray}
which is called the Nernst-Einstein relation. We shall use this expression later on to determine the diffusion constant of probe particles moving in a strongly coupled field theory assuming that the statistical force supplied by fluctuations of the black hole horizon satisfies the conditions necessary for \eqref{Eq:Mobilitydefinition} to apply. See  \cite{deBoer:2008gu} for further discussion.}

\section{Holographic massive gravity}

{\ We use a generalised model of massive gravity to supply the background upon which our string moves. This is inspired by \cite{Baggioli:2014roa} where massive gravity is realized in a fully covariant formalism through the explicit introduction of Stueckelberg fields. This is convenient in order to have easy control on the number of degrees of freedom and possible inconsistency issues for which we refer to \cite{Alberte:2014bua}. The minimal covariantization in the present case requires a set of scalar fields $\Phi^I$  transforming under an internal Euclidean group of translations and rotations in field-space (for further details about this procedure see \cite{Rubakov:2004eb}). Our massive gravity action is then recovered by a truncation to 2-derivative operators:
  \begin{equation}
    \label{Eq:MassiveGravityAction}
    S_{\mathrm{HMG}} = M_{P}^2 \int d^{3+1} x \; \sqrt{-g} \left[\frac{R}{2} + \frac{3}{\ell^2}- \, m^2 V\left( \frac{1}{2} g^{MN} \partial_{M} \Phi^{I} \partial_{N} \Phi^{J} \delta_{IJ} \right)\right] \; ,
  \end{equation}
with $V$ some polynomial. This theory admits solutions where the scalars take on linear expectation values $\bar \Phi^I = \alpha \,\delta^I_i \,x^i$ with $\delta^I_i$ the Kronecker delta\footnote{This choice leaves the Ward identity for energy conservation unbroken while breaking spatial translation invariance.}. On these solutions the mass matrix for the graviton is identified as $\propto\partial_\mu\bar \Phi^I\partial_\nu\bar \Phi^I$.  Note that the Lagrangian for the scalars $\Phi^I$ has some similarities with the effective Lagrangian describing phonons \cite{Leutwyler:1993gf,Nicolis:2013lma}. It is necessary for $V(X)$ to be monotonically increasing in $X$ to avoid ghosts and we choose $V(0)=0$ so that asymptotically AdS spaces are a solution when the scalars vanish.}

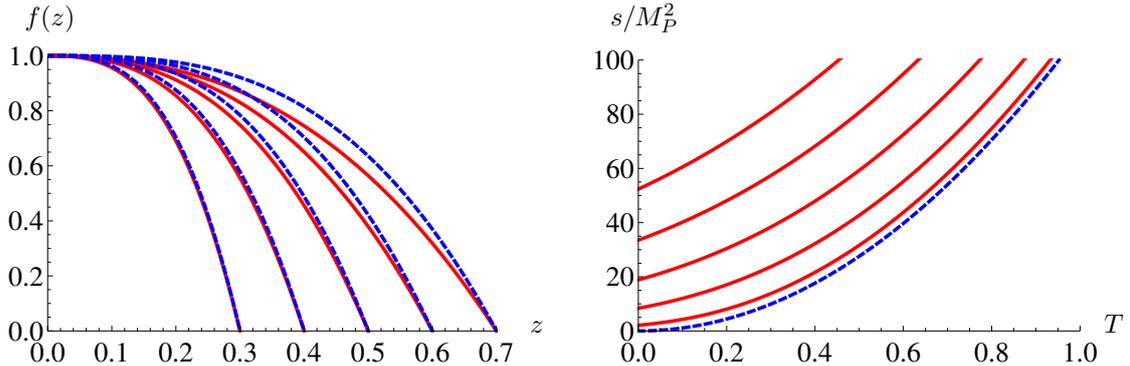
\begin{figure}[t]
 \centering
 \centering
  \begin{subfigure}
  \centering
  \includegraphics[width=0.44\textwidth]{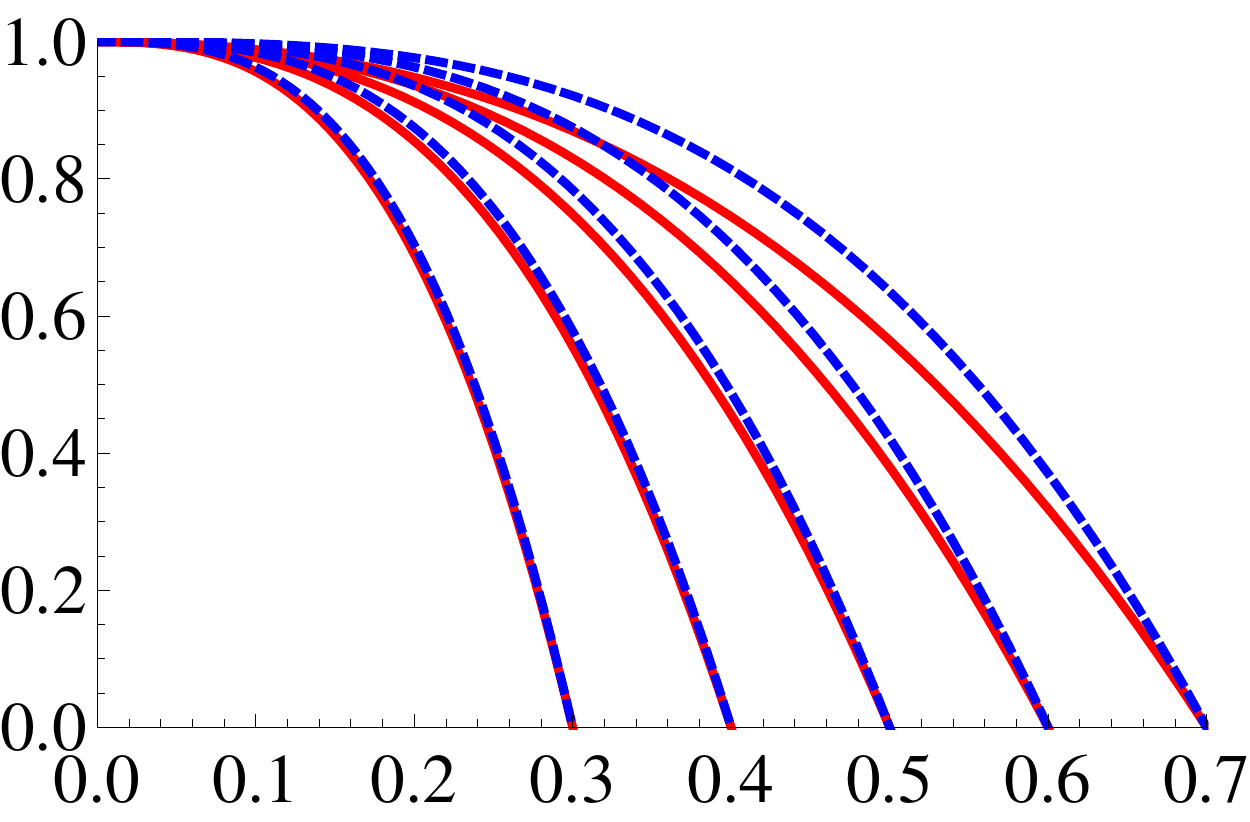}
 \end{subfigure} \qquad
 \begin{subfigure}
  \centering
  \includegraphics[width=0.44\textwidth]{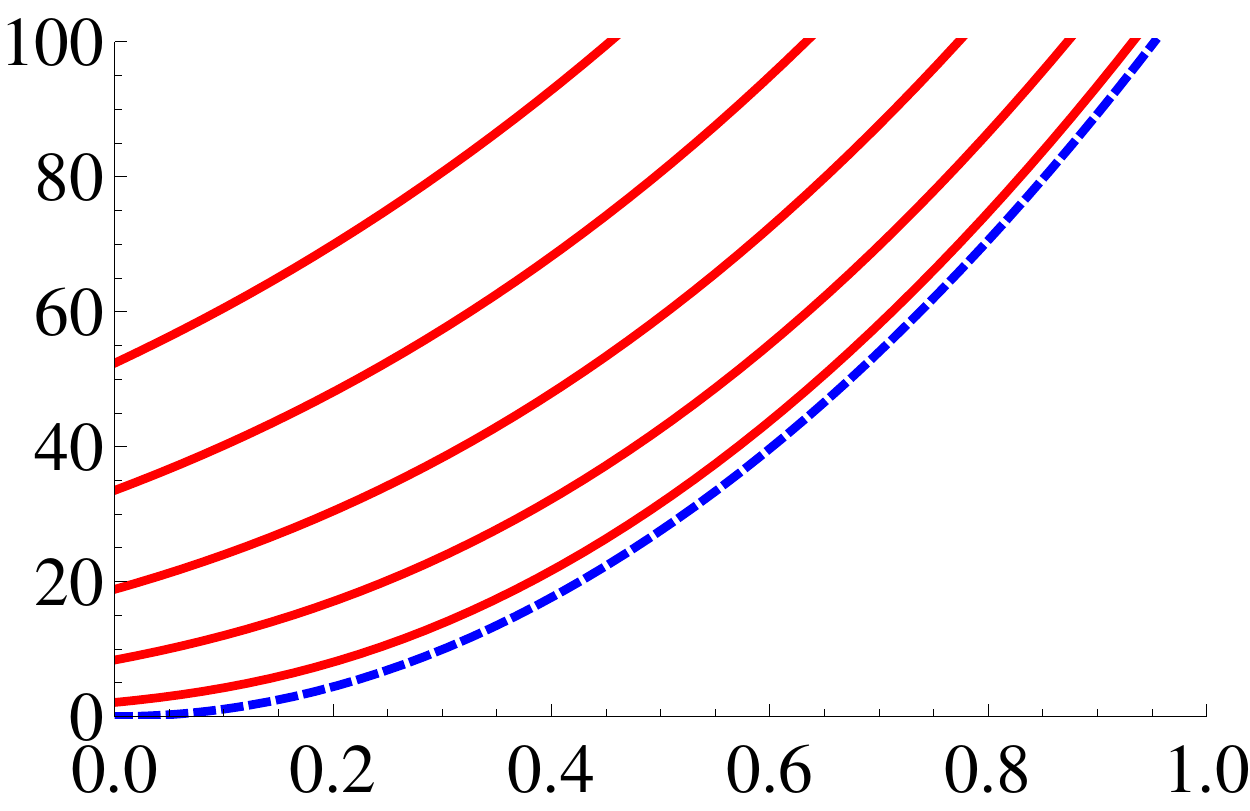}
 \end{subfigure}
 \begin{picture}(100,0)
  \put(4,35){\small{$f(z)$}}
  \put(48,8){\small{$z$}}
  \put(55,35){\small{$s/M_{P}^2$}}
  \put(98,8){\small{$T$}}
 \end{picture}
 \vskip-2em
 \caption{Various plots related to the solution, \eqref{Eq:LineElement} and \eqref{Eq:Emblackeningfactor}, of the background equations of motion. \textbf{Left:} The emblackening factor against radial position in the spacetime for the linear potential $V(X)=X$ at various temperatures. From left to right we set $T \sim 0.77,0.57,0.44,0.35,0.29$ and $m \alpha = 1$. The blue dashed lines represent the zero graviton mass solution at the given temperature while the solid red lines have non-zero graviton mass. \textbf{Right:} The entropy density for vanishing ($V(0)=0$, dashed blue line) and linear potential ($V(X)=X$, red solid lines) against temperature for $m \alpha=1,2,3,4,5$ (bottom solid line to top). Note that for non-zero graviton mass, represented here by the example of the linear potential, the entropy is non-vanishing at zero temperature.}
 \label{fig:EmblackeningandEntropy}
\end{figure}

{\ As the Einstein equations are easy to derive we shall not present them here. A class of metrics which solve these equations are of the form
  \begin{eqnarray}
   \label{Eq:LineElement}
   ds^{2} = g_{MN} \mathrm{d} x^{M} \mathrm{d} x^{N} = \frac{\ell^2}{z^2} \left[ \frac{\mathrm{d}z^2}{f(z)} - f(z) \mathrm{d}t^2 + \mathrm{d}\vec{x}_{2}^2 \right]
  \end{eqnarray}
where $f(z)$, for general potential $V(X)$, reads:
  \begin{eqnarray}
      \label{Eq:Emblackeningfactor}
      f(z) 
    = \left[ 1 - \left( \frac{z}{z_{\mathrm{H}}} \right)^3 \right] - (m \ell)^2 z^3 \int_{w=z}^{z_{\mathrm{H}}} \frac{dw}{w^4} \; V \left( \frac{\alpha^2 w^2}{\ell^2} \right)  \; ,
  \end{eqnarray}
with $\alpha$ being related to some coupling between the scalars. For monomial choices of $V(X)$ one of the parameters, $m$ or $\alpha$, is redundant and can be reabsorbed via an appropiate redefinition. For polynomial $V(X)$ both the parameters are independent and can play a different role. Henceforth we set $\ell=1$.}

{\ Considering \eqref{Eq:Emblackeningfactor} it can be seen that at $z=z_{H}$ we have a coordinate singularity. Despite the known potential issues of singularities in the context of black holes geometries in massive gravity theories \cite{Berezhiani:2011mt}, our solution seems to be totally healthy and smooth. This is because in our scenario the Stueckelberg fields $\phi^I$ have non trivial profiles only in the spatial directions as compared to the Lorentz invariant case discussed in \cite{Berezhiani:2011mt}. Therefore the invariant object $g^{MN}\partial_{M}\Phi^I\,\partial_{N} \Phi^I$ is regular at $z=z_{H}$ and we do not need to concern ourselves with the remedy to these problems described in \cite{Berezhiani:2011mt}. Another quick, but non-comprehensive test, is to consider the standard curvature invariants $R^{2}$, $R_{MN} R^{MN}$ and $R_{MNPQ} R^{MNPQ}$ which all remain finite at $z=z_{H}$.} 

{\ As all the emblackening factors of \eqref{Eq:Emblackeningfactor} have a zero at $z=z_{\mathrm{H}}$, where there is no curvature singularity, our spacetime can be interpreted as thermal with a temperature given by:
  \begin{eqnarray}
     \label{Eq:Temperature}
      T &=& \frac{3}{4 \pi z_{\mathrm{H}}}  \left[ 1 - \frac{m^2}{3}  V\left( \alpha^2 z_{\mathrm{H}}^2 \right) \right] \; .
  \end{eqnarray}
To obtain this temperature we follow the standard procedure of continuing to Euclidean time and imposing smoothness of the manifold at $z=z_{H}$ through suitable periodicity conditions. This ensures that correlation functions on the boundary share this property, the hallmark of a thermal field theory.}

{\ We can see from \eqref{Eq:Temperature} that it is possible to set $T=0$ at non-zero $z_{H}$ and have extremal black holes. It is worth noting that the non-trivial scalars, in particular $\alpha^2$, enter the first law like a magnetic field \cite{Andrade:2013gsa} (or conversely a charge in four spacetime dimensions). As a consequence it is not surprising that one can reach an extremal limit by dialing the graviton mass appropriately as the scalar hair will support the black hole. Indeed assuming that the entropy density is given by the area of the black hole horizon we find
  \begin{eqnarray}
   \label{Eq:EntropyDensity}
   s=2 \pi M_{P}^2/ z_{\mathrm{H}}^2 \; , 
  \end{eqnarray}
which is non-zero at zero temperature if there is a non-trivial solution to \eqref{Eq:Temperature}. The entropy density is also displayed in fig.~\ref{fig:EmblackeningandEntropy} for vanishing graviton mass and an example potential $V(X)=X$. Indeed, for all but a few special cases where the emblackening factor has higher order degeneracies at $z=z_{H}$ the extremal IR geometry turns out to be $AdS_2\,\times \mathbbm{R}^2$; typical of extremal black holes. It would be interesting to study such IR extremal geometries and their instabilities. }

{\ With the above remarks about the thermodynamics in mind we note that the position of the black hole horizon can be given as a function of the temperature and graviton mass by solving
  \begin{eqnarray}
     \label{Eq:Horizonposition}
      T &=& \frac{3}{4 \pi z_{\mathrm{H}}}  \left[ 1 - \frac{m^2}{3}  V\left( \alpha^2 z_{\mathrm{H}}^2 \right) \right] \leq \frac{3}{4 \pi z_{\mathrm{H}}}
  \end{eqnarray}
for $z_{\mathrm{H}}$. This is bounded above by the zero graviton mass result\footnote{To see this assume that both $T$ and $z_{\mathrm{H}}$ must be positive (the possibility of having negative temperature $T<0$ goes beyond the scope of this paper). Thus, when this equation has a solution it must be the case that the quantity in square brackets is positive. Moreover because $V(X)$ is monotonically increasing between zero and $\alpha^2 z_{\mathrm{H}}^2$ (to avoid ghosts) and $V(0)=0$ (so asymptoticlly AdS spacetimes are a solution) the same term is bounded above by one. It equals one only when the second term is zero (for example when the graviton mass is zero). Thus proving the result.}. For the square root ($V(X)=X^{1/2}$) and linear ($V(X)=X$) potentials, upon solving \eqref{Eq:Horizonposition} for $z_{\mathrm{H}}$, we find the simple and exact results:
  \begin{eqnarray}
   \label{Eq:Horizonforpotentials}
   z_{\mathrm{H}} = \frac{3}{4 \pi T} \left\{ \begin{array}{cc}
			 \left( 1 + \frac{m^2  \alpha}{4 \pi T} \right)^{-1} \; , & V(X)=X^{1/2} \\
			 \frac{\sqrt{ 1 + 12 (\frac{m \alpha}{4 \pi T})^2 } - 1}{6 \left( \frac{m \alpha}{4 \pi T} \right)^2} \; , & V(X)=X
                     \end{array} \right. \; .
  \end{eqnarray} 
While there is no such simple result for higher power monomial potentials, generically we see that for larger powers of $X$ in $V(X)$ the leading correction to the zero graviton mass result is smaller. At large temperatures ($m \alpha^{n} \ll T^{1/2}$) we can show that the leading correction to the zero graviton mass horizon position is
  \begin{eqnarray}
     \label{Eq:LargeTHorizonExpansion}
     z_{\mathrm{H}} &=& \frac{3}{4 \pi T} \left[ 1 - \frac{1}{3} \left(\frac{\beta}{T}\right)^{2n} \left( \frac{3}{4 \pi}\right)^{2n} \lim_{X \rightarrow 0} \left( X^{-n} V(X) \right) + \ldots  \right] \; , \\
     \beta &=& m^{1/n} \alpha \; ,
  \end{eqnarray}
where we have assumed the leading term in the potential at small $X$ is proportional to $X^{n}$. Thus if $\beta^{2n} \ll T$ then we can ignore the effect of graviton mass and our spacetime is thermal AdS.}

{\ At low temperatures we face an obstruction to determining $z_{\mathrm{H}}$ in terms of small $T$, namely the massive gravity spacetimes potentially exhibit zero temperature black holes with $z_{\mathrm{H}}$ given by the smallest real solution to
  \begin{eqnarray}
     \label{Eq:ZeroTempHorizonEqn}
     V\left( \alpha^2 z_{\mathrm{H}}^2 \right) &=& \frac{3}{m^2} \; . 
  \end{eqnarray}
These black holes are interesting and particularly worthy of independent study. We note that \eqref{Eq:ZeroTempHorizonEqn} always has a real solution if $V(0)=0$ and $V(X)$ is monotonically increasing. Let $z_{\mathrm{H}}^{(0)}$ be the smallest real solution to \eqref{Eq:ZeroTempHorizonEqn}. For the square root and linear potentials this is given by
  \begin{eqnarray}
   \label{Eq:ZeroTemperatureHorizon}
   z_{\mathrm{H}}^{(0)} &=& \frac{1}{\beta} \left\{
    \begin{array}{ccc}
     3 \; , & V(X)=X^{1/2} \\
    \sqrt{3} \; , & V(X)=X 
    \end{array}
   \right. \;  .
  \end{eqnarray}
For any potential $V(X)$ we find the horizon position as a function of small $T$ is given by
  \begin{eqnarray}
   \label{Eq:SmallTHorizonExpansion}
     z_{\mathrm{H}} &=& z_{\mathrm{H}}^{(0)} \left[ 1 + \frac{1}{\left( 1 + \frac{m^2}{3} z_{\mathrm{H}}^{(0)} V'( z_{\mathrm{H}}^{(0)} ) \right)} T + \mathcal{O}^{2}(T) \right]  \; .
  \end{eqnarray}
The above asymptotic expansions for the horizon position at small and large temperature, \eqref{Eq:SmallTHorizonExpansion} and \eqref{Eq:LargeTHorizonExpansion} respectively, will be particularly useful for determining the behaviour of the drag coefficient and momentum loss rates at the relevant extremes.}

{\ There are of course many possible potentials we could consider here. We will often consider $V(X)=X^{N}$, the monomial potentials\footnote{We have found that the features of generic potentials can often be identified from the features of the monomial potentials. We shall see that the smallest power of $X$ in $V(X)$ is almost always dominant at large temperatures and small $\beta$. In the opposite regime it is often the largest power of $X$ that dominates. The interest in non-linear combinations, such as $V(X)=X+\kappa X^{N}$ comes from the fact that they give rise to metal-insulator transitions \cite{Baggioli:2014roa}.}, with $N=1/2,1,2,3$. The choices $N=1/2$ and $N=1$ are due to the existence of the exact and simple results given in \eqref{Eq:Horizonforpotentials} and \eqref{Eq:ZeroTemperatureHorizon}. For larger $N>4$ we are forced to use numerics exclusively but will also use these techniques for $N=2,3$ as the analytic results are not concise. For intuitive purposes we will consider $V(X)=X+ \kappa 
X^3$, 
with $\kappa$ some positive real number, where appropriate.}

\subsection{The heat capacity}

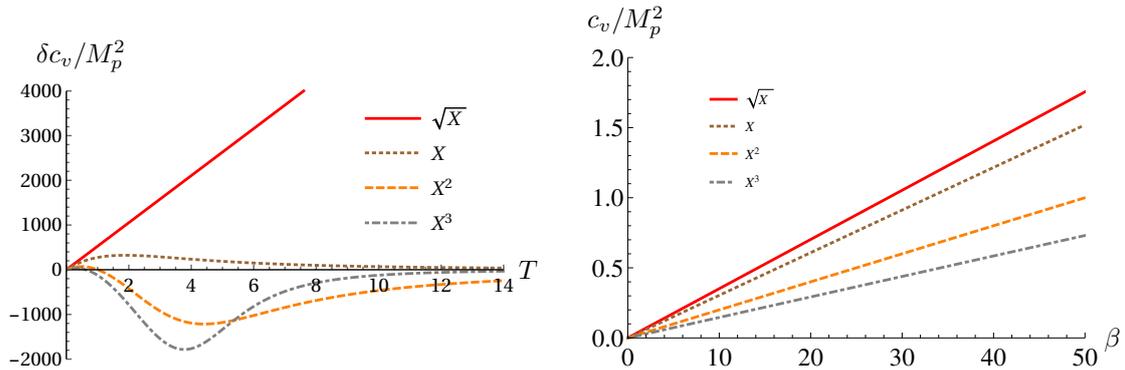
\begin{figure}[t]
 \centering
  \begin{subfigure}
  \centering
  \includegraphics[width=0.44\textwidth]{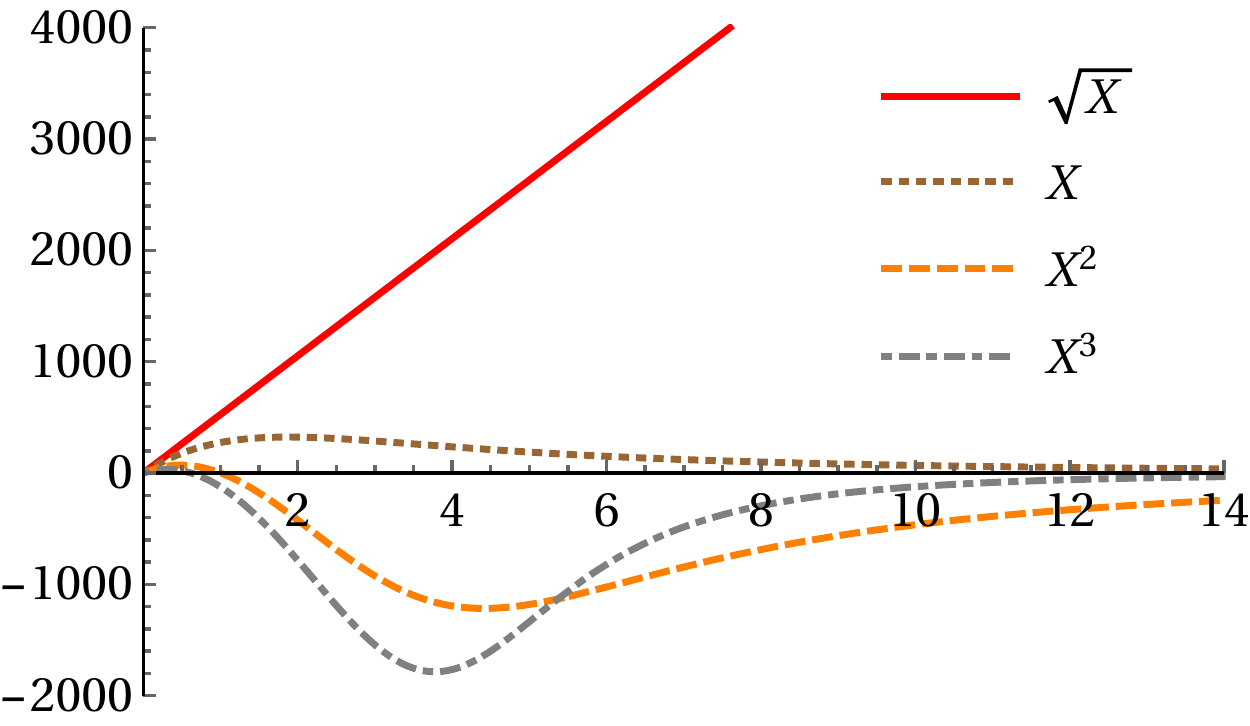}
 \end{subfigure} \qquad
 \begin{subfigure}
  \centering
  \includegraphics[width=0.44\textwidth]{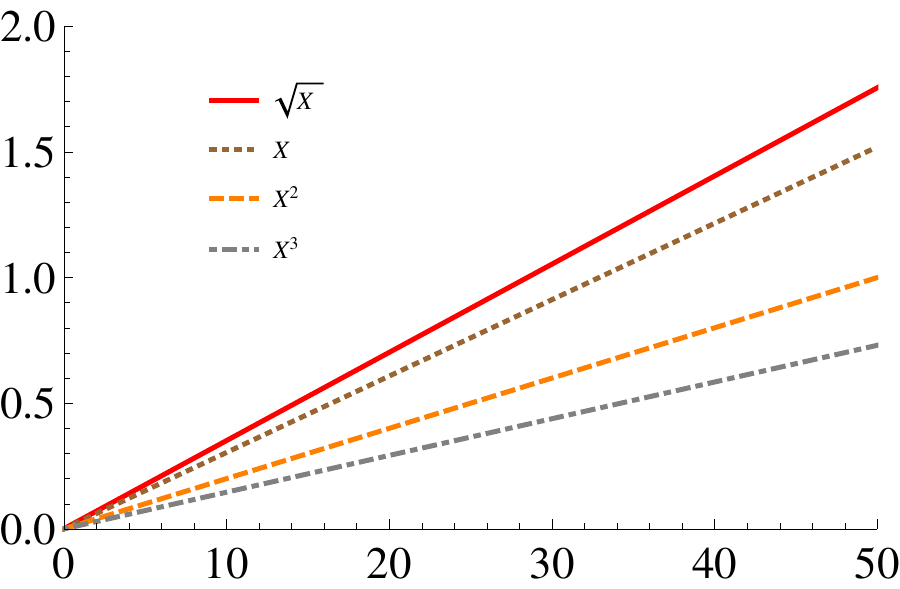}
 \end{subfigure}
 \begin{picture}(100,0)
  \put(5,32){\small{$\delta c_{v}/M_{p}^2$}}
  \put(47,13){\small{$T$}}
  \put(53,35){\small{$c_{v}/M_{p}^2$}}
  \put(98,7){\small{$\beta$}}
 \end{picture}
 \vskip-2em
 \caption{Various plots describing the behaviour of the heat capacity for different potentials. The potentials displayed are the square root ($V(X)=\sqrt{X}$, solid red), linear ($V(X)=X$, dotted brown), quadratic ($V(X)=X^2$, dashed orange) and cubic ($V(X)=X^3$, dot-dashed grey). \textbf{Left:} The change in the heat capacity from the zero graviton mass value against temperature for various potentials and $\beta=15$. The solid line which is always increasing represents the square root potential while the solid line (grey) with the turning point represents the cubic potential. \textbf{Right:} The heat capacity at low temperatures ($T = 10^{-3}$) as a function of $\beta$ for various monomial potentials.}
 \label{fig:HeatCapacity}
\end{figure}

{\ For fixed potentials there are two thermodynamic state variables in our background - volume and temperature. We expect the thermodynamic potential to be extensive in volume and thus we can drop the former leaving only a single scale - the temperature - to be concerned with. The response of the system to thermodynamic forces is encoded in the susceptibilities and for us in particular the volumetric heat capacity at constant volume,
  \begin{equation}
    c_{v} = T \frac{ds}{dT} = \frac{2 (4 \pi)^{2} T M_{P}^{2}}{3 z_{\mathrm{H}}} 
	     \left[ 1 - \frac{m^2}{3} \left( V(\alpha^2 z_{\mathrm{H}}^2) - 2 \alpha^2 z_{\mathrm{H}}^2 V'(\alpha^2 z_{\mathrm{H}}^2) \right) \right]^{-1} \; ,
  \end{equation}
where $s$ is the entropy density of the system defined in \eqref{Eq:EntropyDensity}. We note that for thermal AdS, i.e.~$V(X)=0$, the heat capacity at constant volume is
  \begin{equation}
    \label{Eq:HeatCapacityZeroGraviton}
    \left. c_{v} \right|_{m=0} = \frac{2 (4 \pi)^3 T^2}{3^2} M_{P}^2 \; . 
  \end{equation}
In a typical material there can be several contributions to the heat capacity coming from distinct fields, for example, charged impurities\footnote{Generically uncharged impurities do not contribute to the heat capacity.} and phonon excitations. We now examine the behaviour of this quantity for various potentials to ensure the thermodynamic stability of our background.}

{\ Consider the monomial potentials $V(X)=X^N$. The heat capacities are displayed in fig.~\ref{fig:HeatCapacity}. We can see that for the zero graviton mass result, \eqref{Eq:HeatCapacityZeroGraviton}, as the temperature goes to zero the heat capacity vanishes. This remains true for non-zero graviton mass (left-hand plot of fig.~\ref{fig:HeatCapacity}) and the heat capacity is positive and $\sim T$ at low temperatures. Thus the system, governed by the exotic ground states with non-zero entropy density, is at least locally stable to thermodynamic fluctuations at low temperatures \cite{Davison:2014lua}. This does not guarantee of course that it is stable to other types of fluctuations nor that we have a global thermodynamic minimum. See \cite{Cai:2014znn} for a general discussion of the thermodynamics of black holes in massive gravity.}

{\ From the left plot of fig.~\ref{fig:HeatCapacity} we can see that for the linear potential there is something like a saturation temperature where the effect of the graviton mass is largest. It may be tempting to interpret this as a Debye temperature. However, we note that while calculating the difference between the heat capacity at zero and non-zero graviton mass certainly captures the change in this quantity due to spatial translation symmetry breaking it is difficult to argue that the additional heat capacity is due to the introduction of a new species, such as phonons, alone.}

{\ Higher monomial and polynomial potentials have heat capacities that are reduced from the vanishing graviton mass value over some intermediate range of temperature (fig.~\ref{fig:HeatCapacity} bottom right and top left). Na\"{i}vely one would expect the addition of extra modes to increase heat capacity. This result suggests that some of the entities that contribute to the zero graviton mass heat capacity require more energy to excite as the graviton mass is increased. This would offset the introduction of more modes and lead to the reduced heat capacity observed.}

{\ For square root and linear potentials the heat capacity can be expressed analytically as
  \begin{eqnarray}
    c_{v} = \left. c_{v} \right|_{m=0} \left\{ \begin{array}{cc}
     1 + \frac{\beta}{4 \pi T} \; , & V(X)=X^{1/2} \\
    \frac{1 + 6 \left(\frac{\beta}{4 \pi T}\right)^2 + \sqrt{1+12 \left(\frac{\beta}{4 \pi T}\right)^2}}{2 \sqrt{1+ 12 \left(\frac{\beta}{4 \pi T}\right)^2}} \; , & V(X) = X
    \end{array} \right. \; . 
  \end{eqnarray}
At large temperatures compared to $\beta$ we find the difference in the heat capacity from its zero graviton mass value \eqref{Eq:HeatCapacityZeroGraviton} is
  \begin{eqnarray}
       \delta c_{v}
   &=& - 4 (n-1) M_{P}^2 \beta^{2n} T^{-2(n-1)} \left( \frac{3}{4 \pi}\right)^{2n-3} \lim_{X \rightarrow 0} \left( X^{-n} V(X) \right) + \ldots 
  \end{eqnarray}
except when $n=1$ where the next to subleading term must be considered. It is clear that for $n<1$ at large temperature the change in the heat capacity increases with increasing temperature. When $n>1$ however the opposite is true. This can clearly be seen in the left plot of fig.~\ref{fig:HeatCapacity}. This expression confirms for large temperatures that the correction to the zero graviton mass result is negative when $n>1$ and positive when $1/2<n<1$.}

{\ Importantly, for the monomial potentials we consider and $V(X)=X+\kappa X^3$, with appropriate choices of $\kappa$, the total heat capacity is positive and grows with temperature in the range of temperatures and $\beta$ examined. Thus our background is at least locally stable to thermal fluctuations.}

\section{Motion of a string in holographic massive gravity}

\begin{figure}[t]
 \centering
 \includegraphics[width=0.45\textwidth]{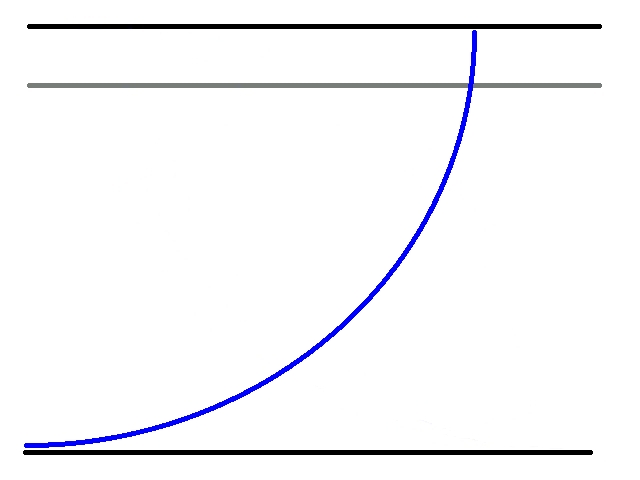}
 \begin{picture}(100,0)
  \put(73,19.5){\vector(0,-10){10}}
  \put(75,14.5){\small{$z$}}
  \put(36,2){\small{Black hole horizon - $z=z_{\mathrm{H}}$}}
  \put(38.5,37){\small{AdS boundary - $z=0$}}
  \put(64,29){\vector(4,0){4}}
  \put(65,27){\small{$v$}}
  \put(30,32){\small{$z=z_{\mathrm{m}}$}}
 \end{picture}
 \vskip-1em
 \caption{A representation of a string stretching from the horizon to the boundary and moving at velocity $v$. The boundary of AdS sits at $z=0$ and the black hole horizon at $z=z_{\mathrm{H}}$. A string stretching to the hyperplane $z=z_{\mathrm{m}}$ from the horizon is given by truncating one that stretches all the way to the boundary. The string drags behind the ``free'' endpoint and falls into the black hole horizon at $z_{\mathrm{H}}$.}
 \label{Fig:stringembedding}
\end{figure}
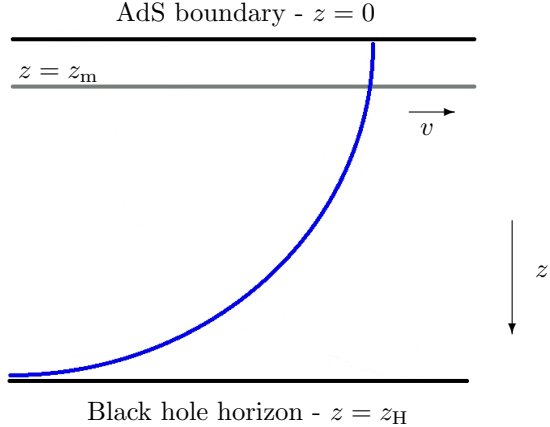

{\ Now that we have discussed the background we turn to calculating the motion of a string in it. We wish to consider a string dangling into the black hole horizon with the other end fixed to a constant $z=z_{\mathrm{m}}$ hyperplace (see fig.~\ref{Fig:stringembedding}). We shall assume that the translation breaking scalars only affect the string through the pullback of the bulk metric to the worldsheet. Thus the string action is just
  \begin{eqnarray}
    \label{Eq:StringAction}
    S = - T_{(0)} \int d^{2} \sigma \; \sqrt{ - \mathrm{det} \left( g_{MN}(X) \partial_{a} X^{M} \partial_{b} X^{N} \right) } \; , 
  \end{eqnarray}
where $T_{(0)}$ is the string tension, $X^{M}(\sigma)$ are the embedding scalars and $\sigma^{a}=(\tau,\sigma)$ are the worldsheet coordinates. The equations of motion for the string take the form
  \begin{eqnarray}
   \label{Eq:GenericStringEquation}
   0 &=& \partial_{\tau} \pi_{\gamma}^{\tau} + \partial_{\sigma} \pi^{\sigma}_{\gamma} 
  \end{eqnarray}
where $\pi_{\gamma}^{\tau}$, $\pi_{\gamma}^{\sigma}$ are given by varying the Lagrangian density with respect to $\partial_{\tau} X^{M}$ and $\partial_{\sigma} X^{M}$ respectively. Assuming the string action takes the form \eqref{Eq:StringAction} may be inconsistent if we attempt to embed \eqref{Eq:MassiveGravityAction} in string theory but as an exercise this allows us to isolate how much the change in the background metric from thermal AdS affects the string and subsequently the drag on the point particle.}

{\ We shall always choose static gauge $\sigma = z$, $\tau = t$ and let the only non-trivial string profile be $x^{1}(z,t)$. The Lagrangian density is
  \begin{eqnarray}
       \label{Eq:UnchargedLagrangian}
    \mathcal{L} = - \frac{T_{(0)} }{z^2} \sqrt{1 + f(z) (\partial_{z} x^{1})^2 - \frac{(\partial_{t} x^{1})^2}{f(z)} } \; , 
  \end{eqnarray}
and the canonical momenta are
  \begin{eqnarray}
       \label{Eq:TauCanonicalMomentum}
       \left( \begin{array}{c}
               \pi_{z}^{\tau} \\ \pi_{t}^{\tau} \\ \pi_{x}^{\tau}
              \end{array} \right)
    &=& \frac{T_{(0)}^2}{z^4 \mathcal{L}} \left( \begin{array}{c}
               \frac{\partial_{t} x^{1}(z,t) \partial_{z} x^{1}(z,t)}{f(z)} \\ 1 + f(z) (\partial_{z} x^{1}(z,t))^2 \\ -\frac{\partial_{t} x^{1}(z,t)}{f(z)}
              \end{array} \right) \; , \\ 
        \label{Eq:SigmaCanonicalMomentum}
        \left( \begin{array}{c}
               \pi_{z}^{\sigma} \\ \pi_{t}^{\sigma} \\ \pi_{x}^{\sigma}
              \end{array} \right)
    &=& \frac{T_{(0)}^2}{z^4 \mathcal{L}} \left( \begin{array}{c}
               1 - \frac{\left(\partial_{t} x^{1}(z,t)\right)^2}{f(z)} \\ - f(z) \partial_{t} x^{1}(z,t) \partial_{z} x^{1}(z,t) \\ f(z) \partial_{z} x^{1}(z,t)
              \end{array} \right) \; . \qquad
  \end{eqnarray}
The equations of motion for the string worldsheet embedding are
  \begin{eqnarray}
   \label{Eq:EmbeddingEquation1}
   &\;& z^4 f(z) \partial_{z} \left( \frac{f(z)}{z^4 \mathcal{L}} \partial_{z} x^{1}(z,t) \right) 
	- \partial_{t} \left(  \frac{\partial_{t} x^{1}(z,t)}{\mathcal{L}} \right) = 0 \; ,
  \end{eqnarray}
with the outgoing boundary condition at $z=z_{\mathrm{H}}$ and where the string is subject a force on the constant $z=z_{\mathrm{m}}$ surface. We note in \cite{Herzog:2006gh} that this force is supplied by a constant electric field on a $D7$-brane whose embedding terminates at $z=z_{\mathrm{m}}$ due to the finite quark mass. Here we shall assume the force can be supplied by some mechanism and remain agnostic about its origin.}

\subsection{The low velocity dispersion relation}
\label{Sec:lowv}

{\ Generically we shall choose the string endpoint not dangling into the horizon to move at small velocities. This allows us to extract analytically the form of the dispersion relation. In this section we shall argue that it takes the non-relativistic form independent of the string length. Firstly we calculate the static energy of the string. Choosing the string to sit at a constant value of $x^{1}(z,t)=x_{\mathrm{static}}$ solves \eqref{Eq:EmbeddingEquation} and gives a total static energy
  \begin{eqnarray}
   \label{Eq:Staticenergy}
    E_{\mathrm{static}}(T) = \int^{z_{\mathrm{H}}}_{z_{\mathrm{m}}} d z \; \pi^{t}_{t}(z,t) = T_{(0)}  \left( \frac{1}{z_{\mathrm{m}}} - \frac{1}{z_{\mathrm{H}}} \right) \; . 
  \end{eqnarray}
The momentum of this static string is zero. While the expression for the static energy \eqref{Eq:Staticenergy} in terms of the horizon position and string endpoint is the same as that found for thermal AdS \cite{Gubser:2006bz,Herzog:2006gh} we note that the dependence of $E_{\mathrm{static}}$ on temperature, through $z_{\mathrm{H}}(\beta,T)$, is different.}

{\ Now consider a string imparted with some initial momentum and allowed to relax. At very late times the drag on the string will make it static. Prior to this point we can consider it as having some small velocity which deforms the string from the static profile. Such an embedding can be described by $x^{1}(z,t) = x_{\mathrm{static}} + \epsilon \delta x^{1}(z,t)$, where $\epsilon$ is a parameter of smallness, and the equation of motion for the fluctuation becomes 
  \begin{eqnarray}
   \label{Eq:Fluctuationeqn}
   z^2 f(z) \partial_{z} \left( \frac{f(z)}{z^2} \partial_{z} \delta x^{1}(z,t) \right) - \partial_{t}^2 \delta x^{1}(z,t) = 0 \; . 
  \end{eqnarray}
Solving for the fluctuation and using \eqref{Eq:TauCanonicalMomentum}, which we do in appendix \ref{sec:mobilityanalytic}, it is possible to determine the dispersion relation at low velocities. For ease we repeat the result,  \eqref{Eq:DispersionRelationAppendix}, here
  \begin{eqnarray}
    \label{Eq:DispersionRelation}
    E = E_{\mathrm{static}}(T) + \frac{\vec{p}^2}{2 M_{\mathrm{eff}}(T)} + \mathcal{O}^{4}(\vec{p}) \; , \qquad \mu^{-1} = \gamma M_{\mathrm{eff}} = \frac{T_{(0)} }{z_{\mathrm{H}}^2} \; ,
  \end{eqnarray}
with $E_{\mathrm{static}}$ given by \eqref{Eq:Staticenergy}. The quantity $\mu$ defined above is the mobility of \eqref{Eq:Mobilitydefinition}. To produce this result we have assumed that the velocity of the string endpoint is given by $\vec{v}(t)=\vec{v}(0) \exp(-\gamma t)$, at late times, with $\gamma$ the as yet undetermined drag coefficient. We shall compute $\gamma$ in section \ref{Sec:QNM} and thus show that this assumption is valid at late times.}

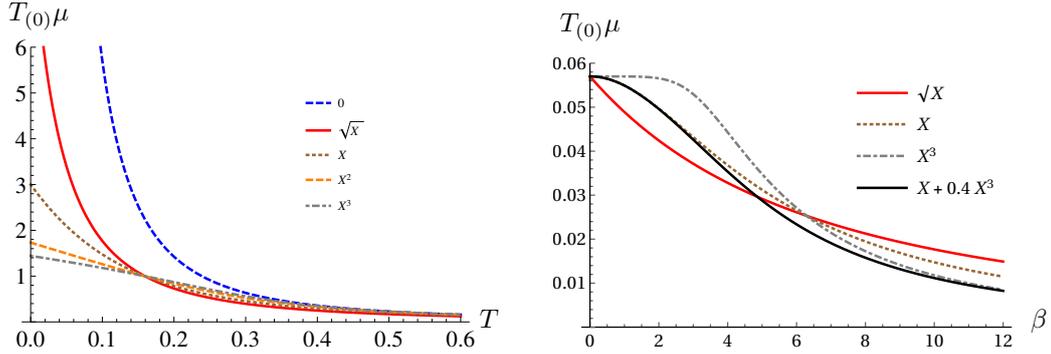
\begin{figure}[t]
 \centering
 \begin{subfigure}
  \centering
  \includegraphics[width=0.4\textwidth]{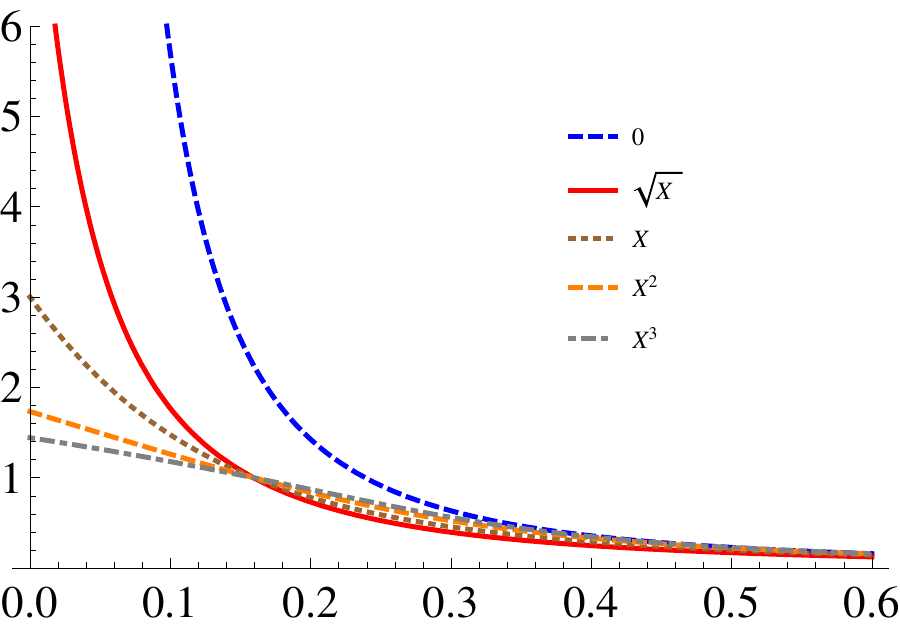}
 \end{subfigure} \qquad
  \begin{subfigure}
  \centering
  \includegraphics[width=0.4\textwidth]{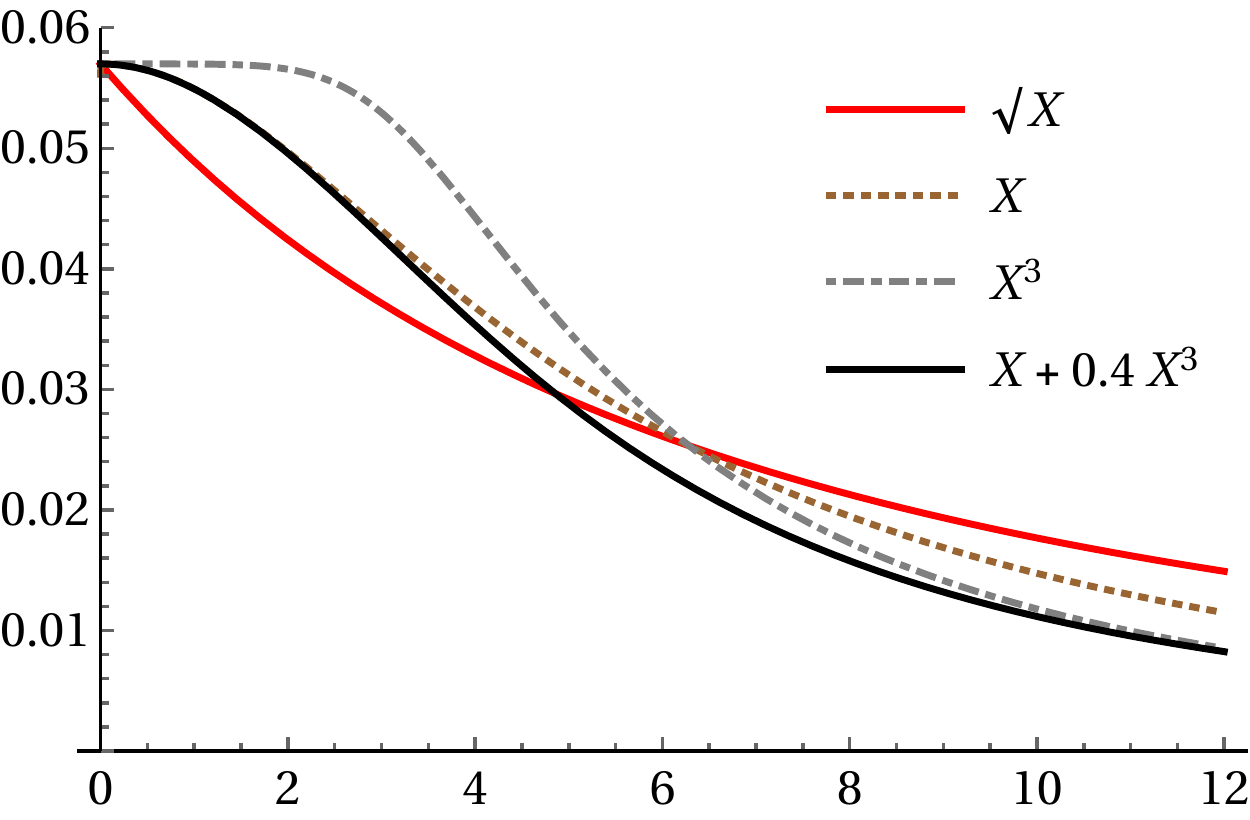}
 \end{subfigure}
 \begin{picture}(100,0)
  \put(6,34){\small{$T_{(0)} \mu$}}
  \put(47,7){\small{$T$}}
  \put(54,33){\small{$T_{(0)} \mu$}}
  \put(95,7){\small{$\beta$}}
 \end{picture}
 \vskip-2em
 \caption{Plots related to the mobility for different potentials against temperature and $\beta$. The potentials displayed are vanishing ($V(X)=0$, dashed blue), square root ($V(X)=\sqrt{X}$, solid red), linear ($V(X)=X$, dotted brown), quadratic ($V(X)=X^2$, dashed orange), cubic ($V(X)=X^3$, dot-dashed grey) and mixed ($V(X)=X+0.4 X^3$, solid black). \textbf{Left:} The mobility against temperature with $\beta=1$. Note that all the potentials at non-zero graviton mass share a crossing point. \textbf{Right:} The mobility against $\beta$ for $T=1$. The mobility for the mixed potential at small $\beta$ is approximately equal to the linear potential while for larger $\beta$ it resembles the cubic potential.}
 \label{fig:Mobility}
\end{figure}

{\ It is important to note before moving on that the mobility of \eqref{Eq:DispersionRelation} is inversely proportional to the entropy density of the field theory. Given that the entropy density increases for non-zero potential (see right plot of fig.~\ref{fig:EmblackeningandEntropy} for example) the decreased mobility of fig.~\ref{fig:Mobility}, for non-vanishing potentials, is consistent with the interpretation of broken translation invariance in the boundary theory. The reason for this is that when we remove the requirement that microstates of the system must preserve spatial translation invariance the number of allowed states increases and there are more channels by which the particle can lose momentum.}
  
{\ Although the graviton mass and potential $V(X)$ enter in the definition of quantities like the entropy and energy there is a single thermodynamic parameter for the background - the temperature - once we fix the potential. Thus we should compare the mobilities in theories with the same temperature normalised by $\beta$.}

{\ Due to the bound on the temperature defined by \eqref{Eq:Horizonposition}, the mobility is bounded above by
  \begin{eqnarray}
   \label{Eq:MobilityUpperBound}
   \mu(T) = \frac{z_{\mathrm{H}}^2}{T_{(0)} } \leq \frac{1}{T_{(0)} } \left( \frac{3}{4 \pi T} \right)^2 = \mu_{\mathrm{bound}}(T) \; . 
  \end{eqnarray} 
The expression for zero graviton mass is equal to $\mu_{\mathrm{bound}}(T)$, displayed in \eqref{Eq:MobilityUpperBound}. For the square root and linear potentials, using \eqref{Eq:Horizonforpotentials}, we find
  \begin{eqnarray}
     \mu(T) &=& \mu_{\mathrm{bound}}(T) \left\{ \begin{array}{cc}
			 \left[ 1 + \frac{\beta}{4 \pi T} \right]^{-2} \; , & V(X)=X^{1/2} \\
			 \left[ \frac{\sqrt{ 1 + 12 (\frac{\beta}{4 \pi T})^2 } - 1}{6 \left( \frac{\beta}{4 \pi T} \right)^2} \right]^2 \; , & V(X)=X
                     \end{array} \right. \; . \qquad
  \end{eqnarray}
These and more general results for the mobility with various choices of the free parameters are displayed against temperature in fig.~\ref{fig:Mobility}. We see that the mobility generically decreases with increasing $\beta$ and decreasing temperature. For all potentials we see that for low values of $\beta$ with respect to $T$, on applying \eqref{Eq:LargeTHorizonExpansion}, 
  \begin{eqnarray}
   \label{Eq:MobilityRatio}
   \frac{\mu(T)}{\mu_{\mathrm{bound}}(T)} =  1 - \frac{2}{3} \left(\frac{\beta}{T}\right)^{2n} \left( \frac{3}{4 \pi}\right)^{2n} \lim_{X \rightarrow 0} \left( X^{-n} V(X) \right) + \ldots 
  \end{eqnarray}
where the leading power of $X$ in $V(X)$ at small $X$ is assumed to be proportional to $X^{n}$. From \eqref{Eq:MobilityRatio} it is also clear that the higher the power of $X$ in $V(X)$ the weaker the correction to the case of zero graviton mass. We have observed that the situation switches at large $\beta$ with respect to $T$. Monotonicity of the potential with these two results is then sufficient to indicate the existence of a value of $z_{\mathrm{H}}$ where any pair of mobilities will have the same value (see left plot of fig.~\ref{fig:Mobility}). For the monomial potentials\footnote{The mobility can be used to partially classify the potentials $V(X)$. Consider $V_{N}(X)=W(X)^{N}$ where $W(X)$ is a polynomial in $X$. The horizon position $z_{\mathrm{H}}$ can be determined by solving
  \begin{eqnarray}
   \frac{m^2}{3} W\left(\alpha^2 z_{H}^2\right)^{N} z_{\mathrm{H}} &=& 4 \pi T z_{\mathrm{H}} - 3 \; . 
  \end{eqnarray}
As we noted earlier, by the monotonicity property of the potential and assuming that at large $\beta$ larger powers of $X$ give greater corrections to the zero graviton mass result, there must be a crossing point for any two potentials. In particular consider $V_{N+1}(X)$ and $V_{N}(X)$. Because the right hand side of the above equation is independent of $N$ it must be the case that
  \begin{eqnarray}
    W\left(\alpha^2 z_{H}^2\right) = \frac{3}{m^2} \; . 
  \end{eqnarray}
The crossing point is given by solving this equation and is independent of $N$. Thus all the potentials of the form $V_{N}(X)=W(X)^{N}$ have a fixed value of the mobility at some value of $z_{\mathrm{H}}$.} this happens when $z_{\mathrm{H}}=\frac{1}{\alpha}$. It would be interesting to understand whether the existence of this point has some physical implications such as enhanced symmetries.}

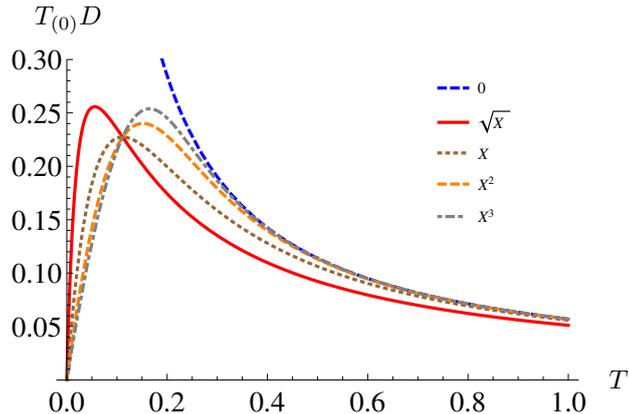
\begin{figure}[t]
 \centering
 \includegraphics[width=0.5\textwidth]{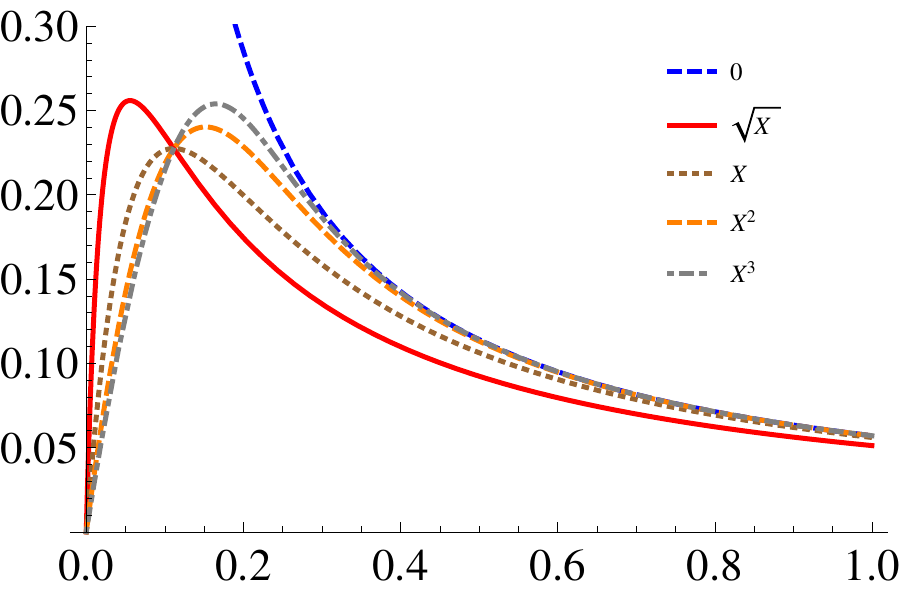}
 \begin{picture}(100,0)
  \put(27,38){\small{$T_{(0)} D$}}
  \put(77,6){\small{$T$}}
 \end{picture}
 \vskip-2em
 \caption{The diffusion constant for various monomial potentials against temperature with $\beta=0.7$. The potentials displayed are vanishing graviton mass ($V(X)=0$, dashed blue line with no turning point), the square root ($V(X)=\sqrt{X}$, solid red), linear ($V(X)=X$, dotted brown), quadratic ($V(X)=X^2$, dashed orange) and cubic ($V(X)=X^3$, dot-dashed grey). Note that for the non-zero mass graviton as the temperature drops to zero the diffusion constant also goes to zero. This is distinct from the zero graviton mass behaviour.}
 \label{Fig:diffusionconstant}
\end{figure}

{\ From \eqref{Eq:Mobilitydefinition} and \eqref{Eq:DispersionRelation} we can additionally determine the diffusion constant for a collection of string endpoints which has the form
  \begin{eqnarray}
    \label{Eq:Diffusionconstant}
    D = T \mu(T) = \frac{z_{\mathrm{H}}^2 T}{T_{(0)} } \; . 
  \end{eqnarray}
Examples of the diffusion constant are displayed in \ref{Fig:diffusionconstant}. Except for the case of vanishing potential there is a peak in the diffusion constant at some value of temperature which moves to larger temperature as the power of the monomial increases. It is also interesting to notice that the diffusion constant goes to zero at $T = 0$ in the case of non-zero graviton mass where the exotic ground state is formed.}

{\ Moreover, because the mobility is bounded above by the zero graviton mass result the diffusion constant is also bounded above by its zero graviton mass value. The existence of this upper bound is deeply related to the fact that $V(X)$ is monotonically increasing (no ghosts) and that $V(0)=0$ (asymptotically AdS spaces are a solution to the background equations of motion). A na\"{i}ve argument of what should happen to the diffusion constant in the presence of broken translation invariance would be that breaking a symmetry permits previously forbidden processes. We expect these to contibute positively to dissipation and thus the diffusion constant to increase. Clearly this cannot be correct alone and the results displayed in fig.~\ref{Fig:diffusionconstant} are consistent with the idea that there is an increased energy gap for processes that contribute to dissipation. This naturally makes the interpretation of our results in terms of new processes, like phonon production or interaction with impurities, more 
difficult 
because the change in the diffusion constant, drag coefficient and so on cannot be attributed to such a process alone.}

{\ Were the reader to have picked up the paper at this point they may be a little surprised as the famous result for thermal transport in holographic systems places a lower bound on the diffusion constant. A bound for diffusion in strongly interacting systems with charge was recently proposed in \cite{Hartnoll:2014lpa} which our system violates badly at sufficiently small temperatures with respect to $\beta$. See \cite{Amoretti:2014ola} for another model which possibly violates this bound. If we are to accept that the conjectured bound of \cite{Hartnoll:2014lpa} is indeed universal this perhaps suggests an instability in our system at low temperatures. To help decide which, if either, solution is correct it is necessary to better understand the interpretation of the string in the field 
theory dual to massive gravity spacetimes.}

{\ We note in passing that many of the dynamical properties of a collection of particles are related to the mobility. It would be interesting to understand whether there are consequences of our bound for other transport quantities such as the thermo-electric transport coefficient recently investigated in \cite{Amoretti:2014zha,Amoretti:2014mma} for duals to massive gravity.}

\subsection{Steady state motion at constant velocity}
\label{Sec:Steady}

{\ For completeness we shall work at arbitrary velocity initially and only take the low velocity limit at the end of the calculation. We consider the following ansatz for the string profile
  \begin{eqnarray}
   \label{Eq:TrailingStringProfile}
   x^{1}(z,t) = v t + \xi(z)
  \end{eqnarray}
which describes a particle moving at constant velocity $v$ in the $x^{1}$ direction on each constant $z$-plane. We substitute \eqref{Eq:TrailingStringProfile} into the equations of motion \eqref{Eq:EmbeddingEquation} and solve for $\xi(z)$ taking the string to stretch from the horizon to the boundary ($z_{\mathrm{m}}=0$). We must impose a force on the string equal to the momentum loss rate to achieve steady state behaviour. Shorter string profiles, where $z_{\mathrm{m}}>0$, can then be found by slicing the profile of the ``full'' string.}

{\ The Lagrangian of \eqref{Eq:UnchargedLagrangian}, assuming \eqref{Eq:TrailingStringProfile}, becomes
  \begin{eqnarray}
    \label{Eq:Lagrangian}
    L = - \frac{T_{(0)} }{z^2} \sqrt{1 + f(z) (d_{z} \xi)^2 - \frac{v^2}{f(z)} } \; . 
  \end{eqnarray}
From this expression it is clear that $\xi$ is a cyclic coordinate so we define
  \begin{eqnarray}
   \label{Eq:CanonicalMomentum}
   \pi_{\xi} = \frac{\delta L}{\delta d_{z} \xi} = - \frac{1}{z^2} \frac{f(z) d_{z} \xi}{\sqrt{1 + f(z) (d_{z} \xi)^2 - \frac{v^2}{f(z)}}} 
  \end{eqnarray}
where $\pi_{\xi}$ is a constant. This expression can be inverted to give a first order differential equation for $\xi$,
  \begin{eqnarray}
   \label{Eq:EmbeddingEquation}
   d_{z} \xi(z) &=& \pm \frac{\pi_{\xi} z^2}{f(z)} \sqrt{\frac{f(z) - v^2}{f(z) - \pi^{2}_{\xi} z^4}} \; ,
  \end{eqnarray}
where the sign is fixed to $-$ by requiring the string to trail behind the free endpoint. We note that because $f(z)$ interpolates monotonically between zero and one and $v^2 \in [0,1]$ there is a value of $z$ where $f(z)-v^2$ changes sign. If the string turns over at this point and returns to the boundary then this is not a problem. However the situation with two string endpoints terminating on the boundary describes a pair of interacting particles. For a single particle the string must past through the black hole horizon and thus the denominator must change sign at the same point (lest the induced metric of the string not be real). As such we must find solutions to the equation
  \begin{eqnarray}
   \label{Eq:TurningPoint}
   f(z_{*}(v)) = v^2 
  \end{eqnarray}
and set $\pi_{\xi}^2  = \frac{v^2}{z_{*}(v)^{4}}$.}

{\ When the spacetime is thermal AdS it is straightforward to solve \eqref{Eq:TurningPoint} to find
  \begin{eqnarray}
     z_{*}(v) = z_{\mathrm{H}} \left( 1 - v^2 \right)^{1/3} \stackrel{v^2 \ll 1}{\approx} z_{\mathrm{H}} \left( 1 - \frac{1}{3} v^2 + \mathcal{O}^{4}(v) \right) \; .
  \end{eqnarray}
However there are corrections for the massive gravity spacetimes and solving \eqref{Eq:TurningPoint} will generically require a numerical approach. In the case of small velocities we can however determine that
  \begin{eqnarray}
   z_{*}(v) = z_{\mathrm{H}} \left[ 1 - \frac{1}{4 \pi T z_{\mathrm{H}}} v^2 + \mathcal{O}^{4}(v) \right] \; , 
  \end{eqnarray}
and thus
  \begin{eqnarray}
    \pi_{\xi}^2  = \frac{v^2}{z_{\mathrm{H}}^{4}} \left( 1 + \frac{1}{\pi T z_{\mathrm{H}}} v^2 + \mathcal{O}^{4}(v) \right) \; . 
  \end{eqnarray}
}

{\ It is sufficient to know $d_{z} \xi(z)$ from solving \eqref{Eq:TurningPoint} to determine the energy and momentum loss rates. Substituting our profile \eqref{Eq:TrailingStringProfile} into \eqref{Eq:CanonicalMomentum} we see that $\partial_{t} \pi^{t}_{\gamma} \equiv 0$. Thus $\pi^{1}_{\gamma}$ is constant along the string following \eqref{Eq:GenericStringEquation}. Hence we can evaluate $\pi^{1}_{\gamma}$ at any point on the string including at the horizon. Using \eqref{Eq:CanonicalMomentum} we can thus determine the loss rates to be
  \begin{eqnarray}
   \pi_{t}^{x} = - T_{(0)} v \pi_{\xi} \; , \qquad \pi_{t}^{x} = - v \pi_{x}^{x} \; ,
  \end{eqnarray}
respectively. In the low velocity limit the momentum loss rate becomes
  \begin{eqnarray}
   \pi_{x}^{x} &=& \frac{T_{(0)} v}{z_{\mathrm{H}}^2} \left[ 1 + \frac{1}{2 \pi T z_{\mathrm{H}}} v^2 + \mathcal{O}^{4}(v) \right] \; . 
  \end{eqnarray}
It is important to now clarify that there is non-zero energy loss for the string even in the case of non-zero graviton mass due to the probe limit. The way that linear sources for our scalars were chosen in \eqref{Eq:MassiveGravityAction} means we are only breaking the subgroup of the full Lorentz symmetry responsible for momentum conservation. The Ward identity for the time component is kept untouched and energy is a completely conserved quantity in the background theory.}

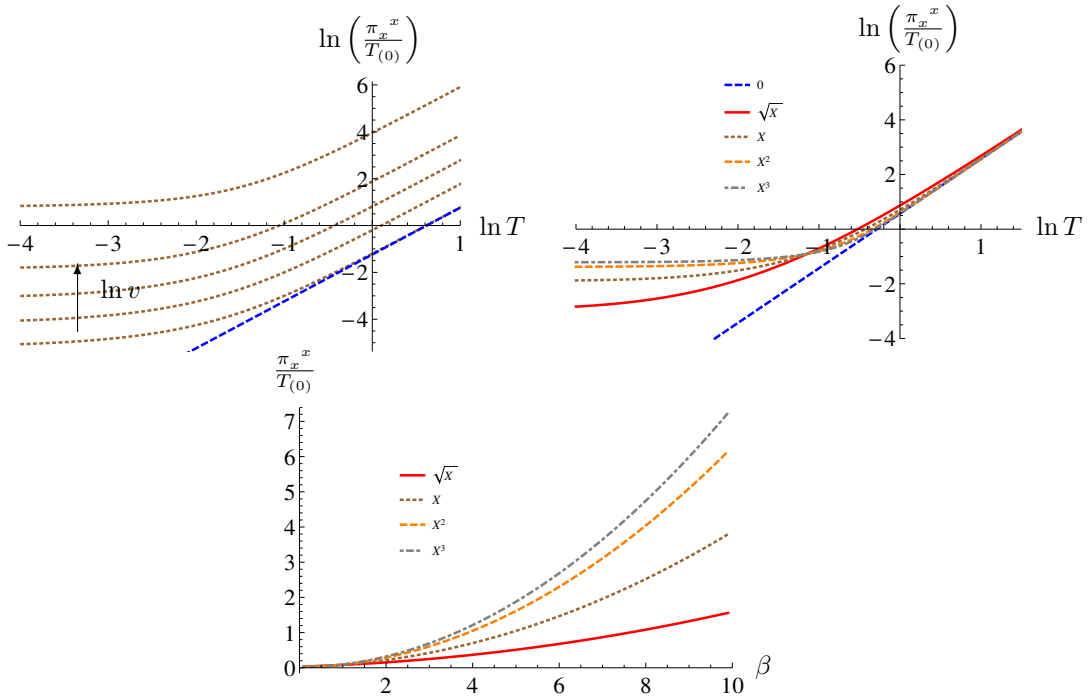
\begin{figure}[t]
 \centering
 \begin{subfigure}
  \centering
  \includegraphics[width=0.4\textwidth]{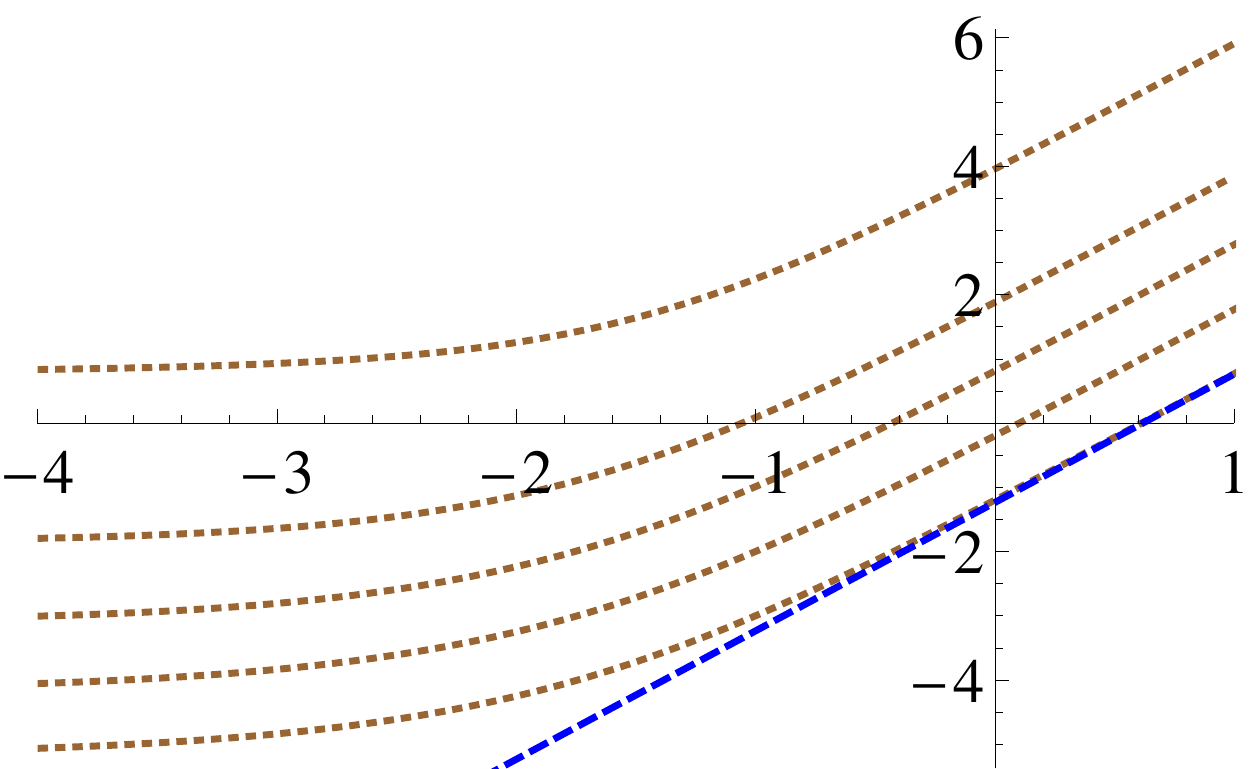}
 \end{subfigure} \qquad \;
 \begin{subfigure}
    \centering
    \includegraphics[width=0.4\textwidth]{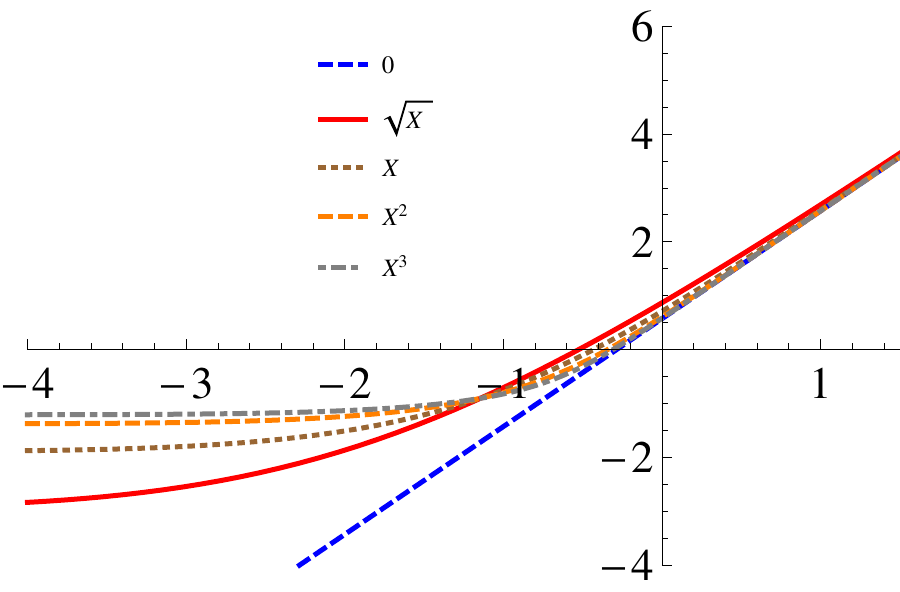}
 \end{subfigure} \\ 
 \begin{subfigure}
  \centering
  \includegraphics[width=0.4\textwidth]{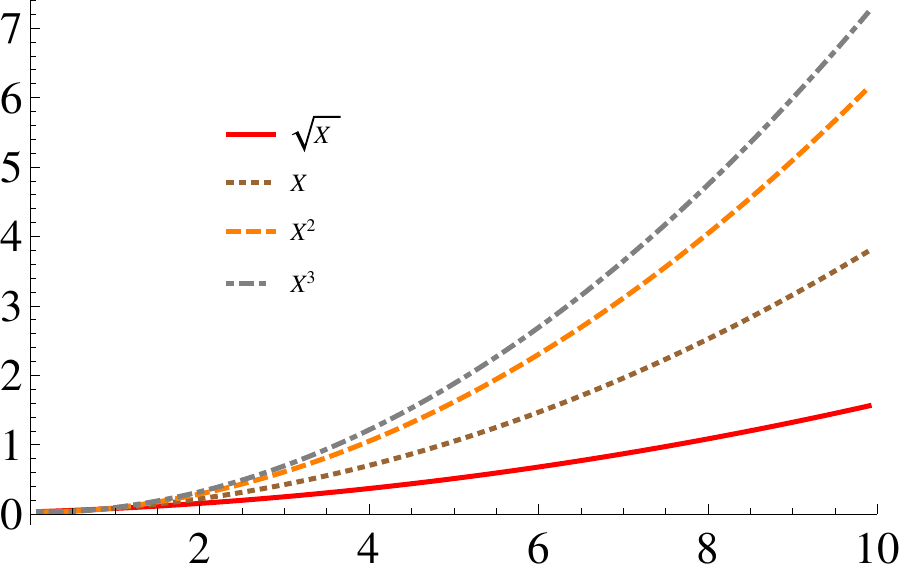}
 \end{subfigure}
 \begin{picture}(100,0)
  \put(33,62){\small{$\ln \left(\frac{\pi\indices{_{x}^{x}}}{T_{(0)}}\right)$}}
  \put(47,45.5){\small{$\ln T$}}
  \put(80,63){\small{$\ln \left(\frac{\pi\indices{_{x}^{x}}}{T_{(0)}}\right)$}}
  \put(95.5,45.5){\small{$\ln T$}}
  \put(29,33){\small{$\frac{\pi\indices{_{x}^{x}}}{T_{(0)}}$}}
  \put(71,7){\small{$\beta$}}
  \put(12,37){\vector(0,6){6}}
  \put(14,40){\small{$\ln v$}}
 \end{picture}
 \vskip-2em
 \caption{Various plots of the momentum loss obtained from several potentials. The potentials displayed are vanishing ($V(X)=0$, straight dashed blue lines), square root ($V(X)=\sqrt{X}$, solid red), linear ($V(X)=X$, dotted brown), quadratic ($V(X)=X^2$, dashed orange) and cubic ($V(X)=X^3$, dot-dashed grey). \textbf{Top left:} The logarithm of the momentum loss rate for the linear potential against the logarithm of temperature at $\beta=1$ and $\ln(v) = -4.1,-3.1,-2.1,-1.1,-0.1$ (bottom dotted line to top dotted line). The blue dashed line represents the zero graviton mass result. \textbf{Top right:} Logarithm of the momentum loss rate against the logarithm of temperature for various potentials with $v=0.1$ and $\beta=2$. The zero graviton mass line (blue dashed) has gradient two in agreement with \eqref{Eq:ZeroMassMomentumLoss} and we see at small temperature the loss rate tends to a constant for all other potentials. \textbf{Bottom:} The momentum loss rate against $\beta$ for monomial 
potentials with $\ln(T)=-2$ and $v=0.1$. It is clear that as the power of the monomial and $\beta$ increase so does the loss rate.}
 \label{fig:Momentumloss}
\end{figure}

{\ For the square root and linear potentials we can determine the momentum loss rate analytically as $f(z_{*}(v)) = v^2$ is at most quadratic in $z_{*}(v)$. Only the linear potential has a compact form however which we display for posterity
  \begin{eqnarray}
   z_{*}(v) &=& z_{\mathrm{H}}(T,\beta) \left[ \frac{(1-v^2) - z_{\mathrm{H}}(T,\beta)^3 \left( \frac{\beta}{3} \right)^2}
		{1 - z_{\mathrm{H}}(T,\beta)^3 \left( \frac{\beta}{3} \right)^2} \right]^{1/3} \; ,  \\
   z_{\mathrm{H}}(T,\beta) &=& \frac{3}{4 \pi T} \left( \frac{\sqrt{ 1 + 12 (\frac{\beta}{4 \pi T})^2 } - 1}{6 \left( \frac{\beta}{4 \pi T} \right)^2} \right) \; .
  \end{eqnarray}
Noting that $v$ enters as a square we see that only for quite relativistic velocities is there a signficant difference from the $v=0$ result. We display the momentum loss for generic potentials in fig.~\ref{fig:Momentumloss}. Beginning with the linear potential in the left plot we see that at large temperatures the momentum loss tends to the same large $T$ behaviour i.e.~$\sim T^2$. This large $T$ behaviour is the same as the momentum loss of the zero graviton mass case,
  \begin{eqnarray}
   \label{Eq:ZeroMassMomentumLoss}
   \pi\indices{_{x}^{x}} \stackrel{T \gg 1}{\sim} \frac{v T^2}{\left( 1 - v^2 \right)^{2/3}} \; , 
  \end{eqnarray}
which is sensible because the effect of graviton mass is reduced by large temperature. At small temperatures however there is a non-vanishing momentum loss for non-zero graviton mass unlike in the case of zero graviton mass. Gravitationally this is because there are non-trivial solutions to \eqref{Eq:ZeroTempHorizonEqn} for the linear potential. In terms of the field theory we can see that the entropy density is non-zero at zero temperature. Thus there are states that the point particle can lose energy and momentum to.}

{\ The other monomial potentials follow the same generic pattern. At large temperatures and small $\beta$ we find that the momentum loss tends to \eqref{Eq:ZeroMassMomentumLoss}. At the opposite extreme we find that the formation of the zero temperature black holes leads to a momentum loss approximately independent of temperature. Physically there is a non-zero density of states for the string to radiate momentum to at zero temperature as indicated by the non-zero entropy density. Again, we can understand this from the gravity side by solving for the position of the horizon at zero temperature
  \begin{eqnarray}
   z_{\mathrm{H}}^{(0)} &=& \frac{3^{1/2N}}{\beta} \; .
  \end{eqnarray}
As the power of the monomial potential increases the momentum loss at small temperatures increases (see top-left of fig.~\ref{fig:Momentumloss}). Finally, from the physical point of view, increasing the power in the monomial potential should increase the strength of the interactions responsible for momentum dissipation and therefore increase the momentum loss itself. This can be seen from the bottom plot in fig.~\ref{fig:Momentumloss}.}

{\ On the condition that the velocity is small the momentum loss has the following generic behaviour
  \begin{eqnarray}
     \label{Eq:MomentumLossApprox}
     \pi_{x}^{x} \sim v s \sim v m^{2/N} \alpha^{2}
  \end{eqnarray}
where again $s$ is the entropy density. This is a lower bound for the momentum loss in the velocity as for non-zero $v$ we see that $z_{*}(v)<z_{\mathrm{H}}$. Unsurprisingly the momentum loss increases as the velocity increases. The approximate relation for the momentum loss \eqref{Eq:MomentumLossApprox} at small velocity also indicates that for $m>1$ as $N$ increases the momentum loss increases while for $m<1$ it decreases. This can be seen in fig.~\ref{fig:Momentumloss} where the zero temperature momentum loss increases with increasing $N$.}

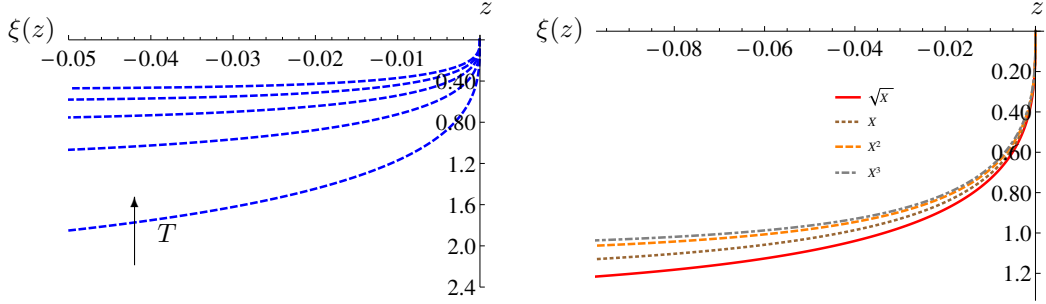
\begin{figure}[t]
 \centering
 \begin{subfigure}
  \centering
  \includegraphics[width=0.4\textwidth]{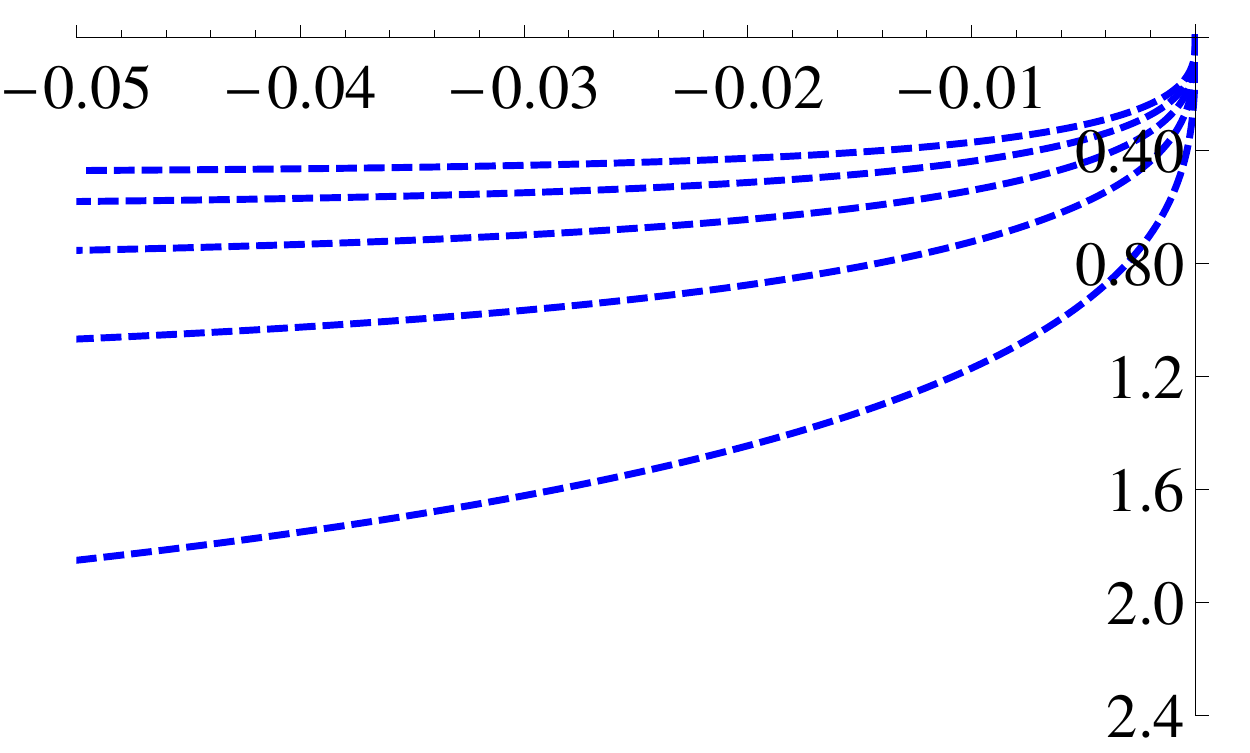}
 \end{subfigure} \qquad \;
 \begin{subfigure}
  \centering
  \includegraphics[width=0.4\textwidth]{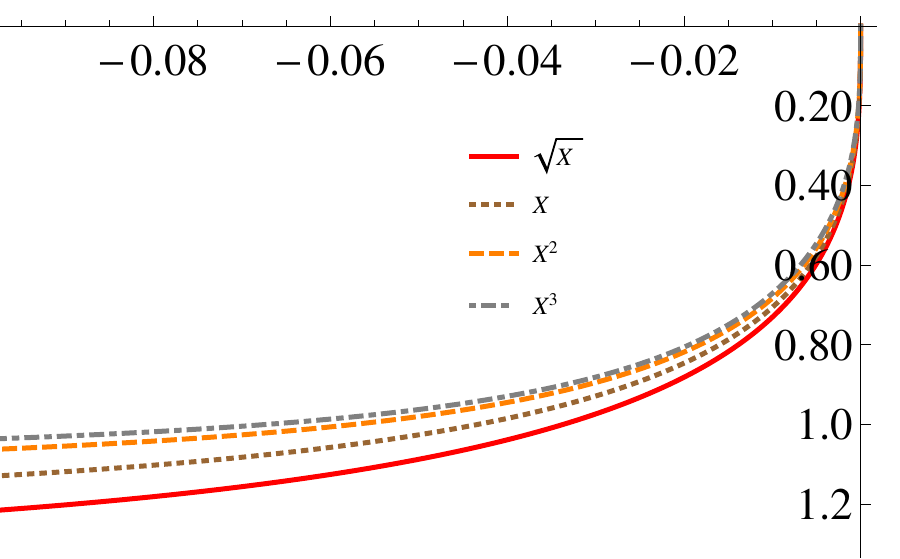}
 \end{subfigure} 
 \begin{picture}(100,0)
  \put(44,30){\small{$\small{z}$}}
  \put(49,28){\small{$\small{\xi(z)}$}}
  \put(92,30){\small{$\small{z}$}}
  \put(3,28){\small{$\small{\xi(z)}$}}
  \put(14,8){\vector(0,6){6}}
  \put(16,10){\small{$\small{T}$}}
 \end{picture}
 \vskip-2em
 \caption{The embedding of the string obtained from several potentials to be compared with fig.~\ref{Fig:stringembedding}. The potentials displayed are vanishing ($V(X)=0$, dashed blue line with no turning point), square root ($V(X)=\sqrt{X}$, solid red), linear ($V(X)=X$, dotted brown), quadratic ($V(X)=X^2$, dashed orange) and cubic ($V(X)=X^3$, dot-dashed grey). \textbf{Top left:} The embedding of the string for vanishing graviton mass with $v=0.1$ and temperatures $T=0.1,0.2,0.3,0.4,0.5$ (bottom to top respectively). As the string asymptotes to the horizon position at $z=z_{\mathrm{H}}$ the $x^{1}$ coordinate diverges. \textbf{Top right:} The string embedding for various monomial potentials with $\beta=1$, $T=1/10$ and $v=1/10$. For the embeddings to be distinguishable we have chosen $\beta\gg T$.}
 \label{fig:actualstringembedding}
\end{figure}

{\ Having computed the value of $\pi_{\xi}$ the embedding equation \eqref{Eq:EmbeddingEquation} must generically be numerically integrated to determine $\xi(z)$. At small $z$ we see from \eqref{Eq:EmbeddingEquation} that 
  \begin{eqnarray}
   d_{z} \xi(z) &=& - z^2 \pi_{\xi} \sqrt{1 - v^2} + \ldots 
  \end{eqnarray}
and thus we require that $\xi(0)$ is a constant. Subsequently we can integrate \eqref{Eq:EmbeddingEquation} to determine the string profile 
  \begin{eqnarray}
   \label{Eq:StringProfile}
   x^{1}(z,t) &=&  v t - | \pi_{\xi}(v) | \int_{s=0}^{z} ds \frac{s^2}{f(s)} \sqrt{\frac{f(s) - v^2}{f(s) - \pi^{2}_{\xi}(v) s^4}} \; ,
  \end{eqnarray}
which we display in fig.~\ref{fig:actualstringembedding}. To describe a string stretching from a constant $z$-surface, $z=z_{\mathrm{m}}$, to the horizon that starts at $x^{1}=0$ at $t=0$ we simply slice our embedding and use the coordinate shift
  \begin{eqnarray}
   x^{1} \rightarrow x^{1} + | \pi_{\xi}(v) | \int_{s=0}^{z_{\mathrm{m}}} ds \frac{s^2}{f(s)} \sqrt{\frac{f(s) - v^2}{f(s) - \pi^{2}_{\xi}(v) s^4}} \; . 
  \end{eqnarray}
}

{\ As we noted the canonical momenta along the string do not vanish. Following \cite{Gubser:2006bz,Herzog:2006gh} a force must be applied to the string endpoint at $z=z_{\mathrm{m}}$ so that we cancel the momentum loss and subsequently satisfy the variational principle. Setting the force equal to the momentum loss rate and substituting into the force balance equation yields
  \begin{eqnarray}
   \vec{f} - \gamma \vec{p} = 0 \qquad \Rightarrow \qquad 
   \gamma M_{\mathrm{eff}} = \frac{\vec{f} \cdot \vec{v}}{\vec{v}^2} = \frac{T_{(0)} }{z_{\mathrm{H}}^2} \left[ 1 + \frac{1}{2 \pi T z_{\mathrm{H}}} \vec{v}^2 + \mathcal{O}^{4}(v) \right] \; ,
  \end{eqnarray}
where here $\cdot$ indicates the usual dot product and we have assumed low velocities so that $\vec{p}=M_{\mathrm{eff}} \vec{v}$. We see that the expression for $M_{\mathrm{eff}}$ to the lowest order in $\vec{v}^2$ matches that given in \eqref{Eq:DispersionRelation}.}

\subsection{Late time unforced motion}
\label{Sec:QNM}

{\ Earlier, in section \ref{Sec:lowv}, we described the behaviour of a particle at late times but neglected to determine the decay constant $\gamma$ as this required a numerical quasi-normal mode analysis. We now seek to remedy this omission. Using the Fourier decomposition of \eqref{Eq:Fourierfluc} and substituting into \eqref{Eq:Fluctuationeqn} leads to
  \begin{eqnarray}
    \label{Eq:QNMequation}
    z^2 f(z) \partial_{z} \left( \frac{f(z)}{z^2} \partial_{z} \delta x^{1}(z,t) \right) + \omega^2 \delta x^{1}(z,t) = 0 \; .  
  \end{eqnarray}
We will solve this equation numerically for the lowest quasi-normal mode as a function of the temperature and $\beta$ by imposing outgoing conditions on the past horizon and then searching for $\omega$ such that $d_{z} \delta x^{1}(z,t) \equiv 0$ at some choice of $z_{\mathrm{m}}$. There are also two regimes where we can find analytic results to compare against these numerics: small and large static energies.}

{\ Before performing this analysis we note that our major results so far (namely the mobility and momentum loss rates) in sections \ref{Sec:lowv} and \ref{Sec:Steady} were independent of the string length. However the drag coefficient is dependent on this quantity. We have already argued we should compare theories of massive gravity at the same temperature. Now we make the choice to compare strings in those theories with the same static energy \eqref{Eq:Staticenergy}. For a choice of mass and potential picking the temperature amounts to fixing $z_{\mathrm{H}}$. Comparing strings of the same static energy then fixes $z_{\mathrm{m}}$. Thus the free parameters are the temperature $T$, static energy $E_{\mathrm{static}}$, $\beta$ and the potential.}

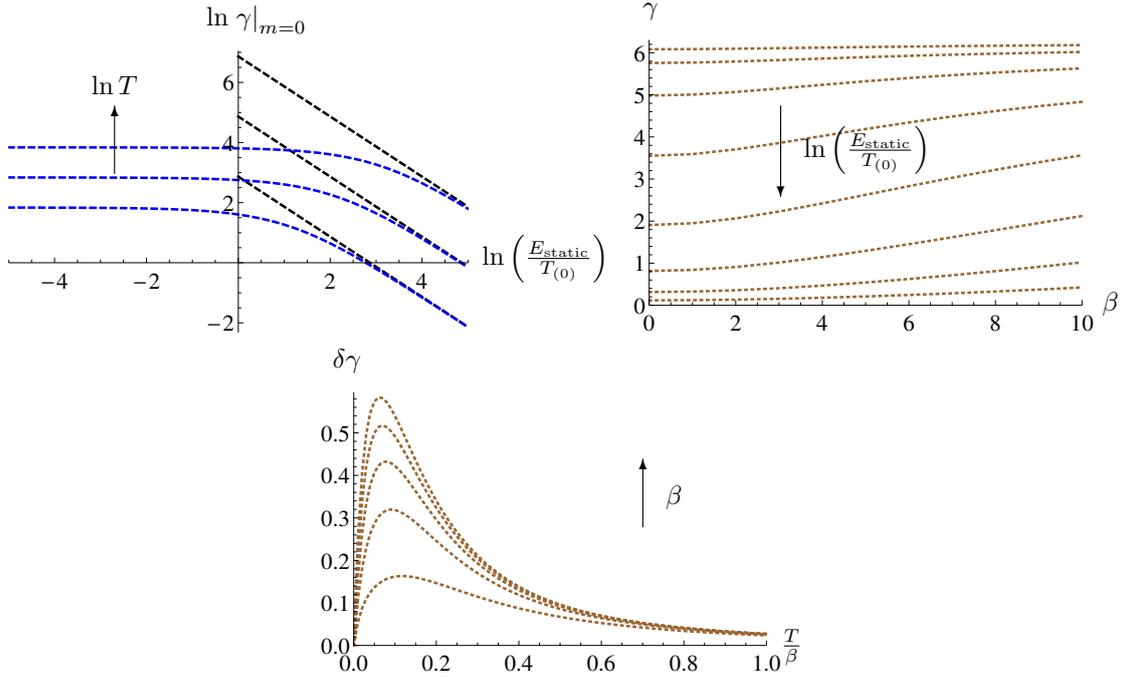
\begin{figure}[t]
 \centering
 \begin{subfigure}
  \centering
  \includegraphics[width=0.4\textwidth]{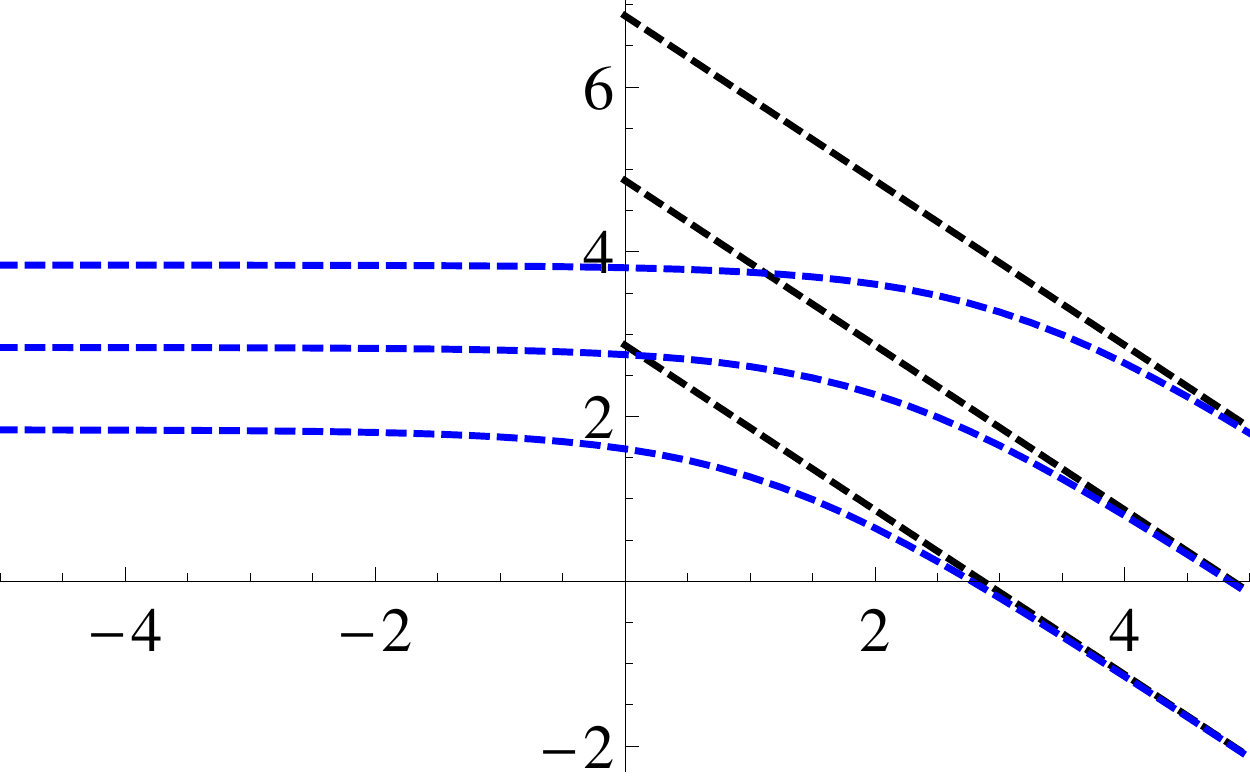}
 \end{subfigure} \hskip+5em
 \begin{subfigure}
  \centering
  \includegraphics[width=0.4\textwidth]{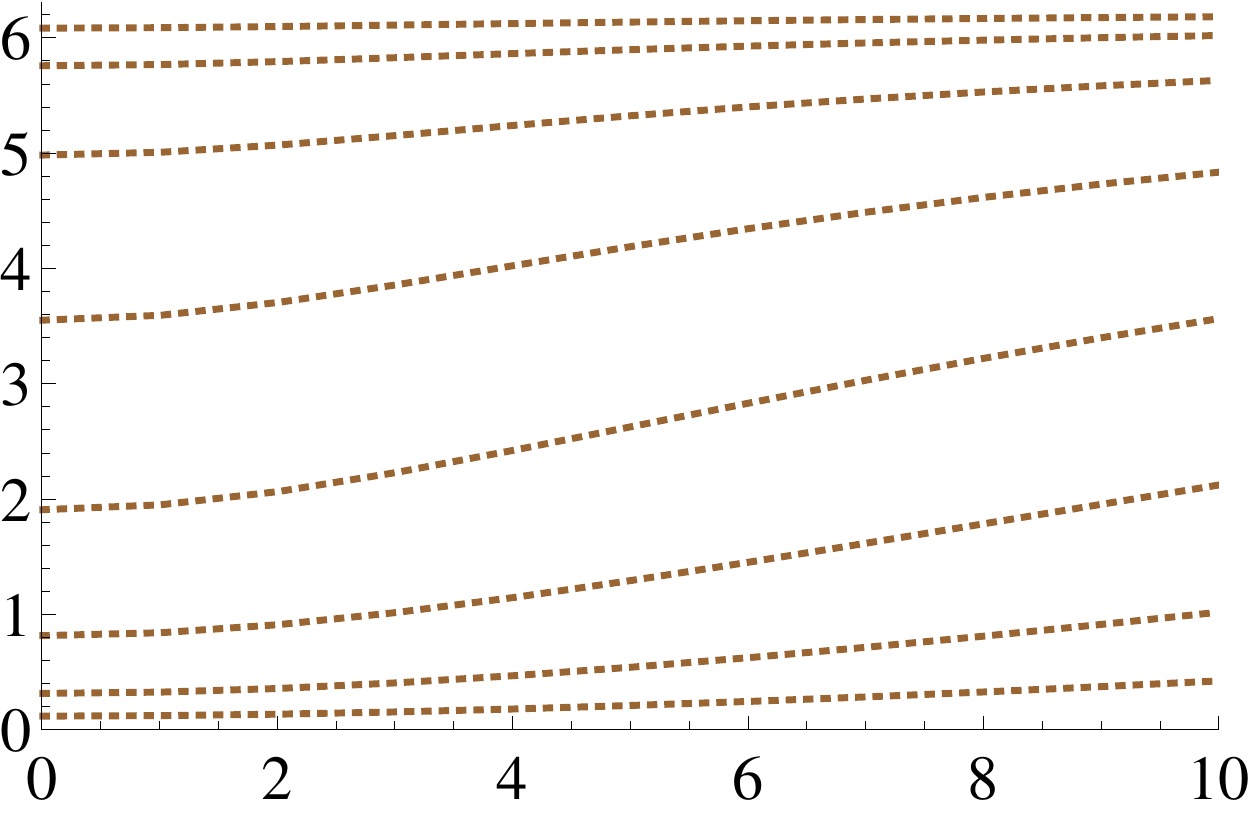}
 \end{subfigure} \\
 \begin{subfigure}
 \centering
  \includegraphics[width=0.4\textwidth]{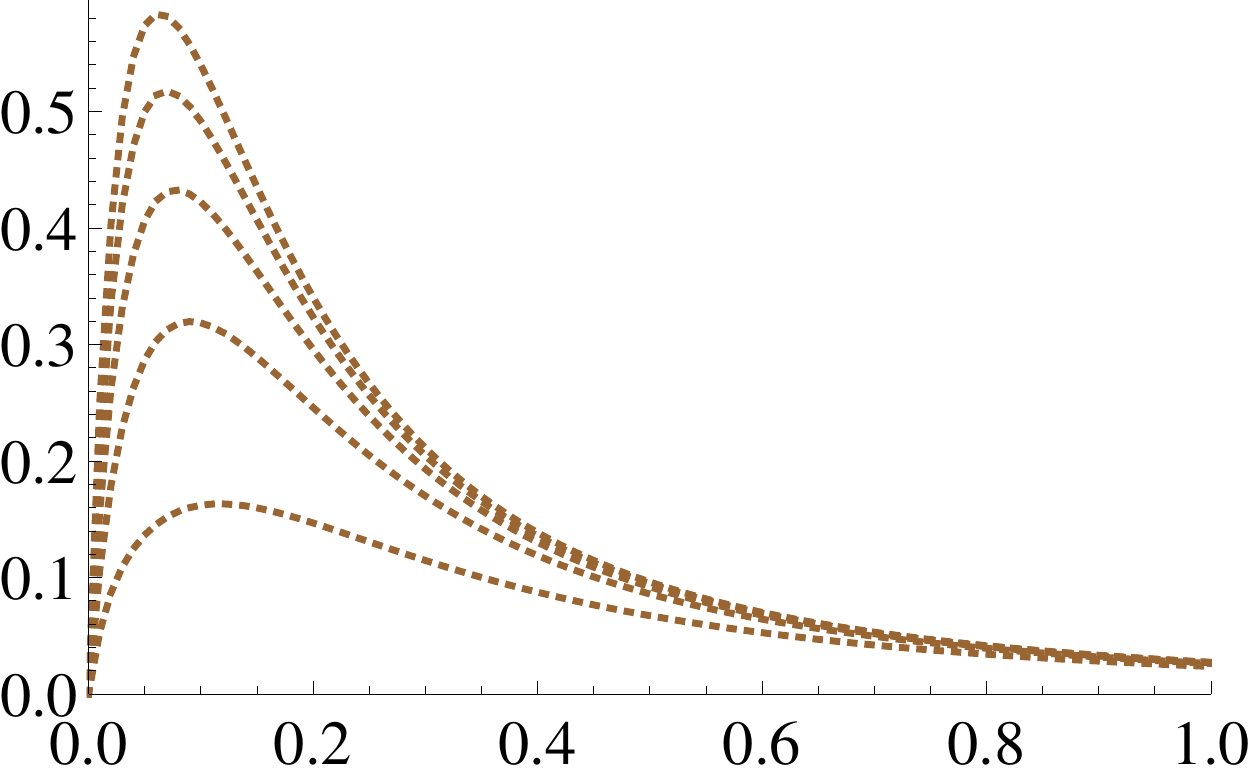}
 \end{subfigure} 
 \begin{picture}(100,0)
  \put(20,62){\small{$\ln \left. \gamma \right|_{m=0}$}}
  \put(44,41){\small{$\ln \left(\frac{E_{\mathrm{static}}}{T_{(0)}} \right)$}}
  \put(10,56){\small{$\ln T$}}
  \put(12,49){\vector(0,6){6}}
  \put(58,63){\small{$\gamma$}}
  \put(98,37){\small{$\beta$}}
  \put(72,50.5){\small{$\ln \left(\frac{E_{\mathrm{static}}}{T_{(0)}}\right)$}}
  \put(70,55){\vector(0,-8){8}}
  \put(31,32){\small{$\delta \gamma$}}
  \put(60,20){\small{$\beta$}}
  \put(70,7){\small{$\frac{T}{\beta}$}}
  \put(58,18){\vector(0,6){6}}
 \end{picture}
 \vskip-2em
 \caption{Various plots related to the drag coefficients for vanishing graviton mass ($V(X)=0$ dashed blue) and linear potential ($V(X)=X$, dotted brown). \textbf{Top left:} The logarithm of the drag coefficient at zero graviton mass against $\ln E_{\mathrm{static}}$ for $\ln T=0,1,2$ (lowest to highest dashed blue lines). The numerical results are given by the dashed blue curves and tend to a constant at small $E_{\mathrm{static}}$ spaced linearly in $\ln T$. The black straight dashed lines are given by the analytic results in \eqref{Eq:DragZeroMassLargeEstatic} and are thus separated vertically by $2 \ln T$. \textbf{Top right:} The drag coefficient against $\beta$ for the linear potential with $\ln E_{\mathrm{static}}=-2,-1,\ldots,4,5$ (top to bottom line) and $T=1$. We see that as $E_{\mathrm{static}}$ becomes small or large the drag coefficient becomes approximately independent of $\beta$. \textbf{Bottom:} The difference between the drag coefficient for the linear potential and that at zero graviton 
mass against temperature. We fix $E_{\mathrm{static}}/T_{(0)}=1$ and choose $\beta = 1,2,3,4,5$ (bottom to top line respectively).}
\label{fig:DragCoefficients}
\end{figure}

\begin{figure}[t]
 \centering
 \begin{subfigure}
  \centering
  \includegraphics[width=0.4\textwidth]{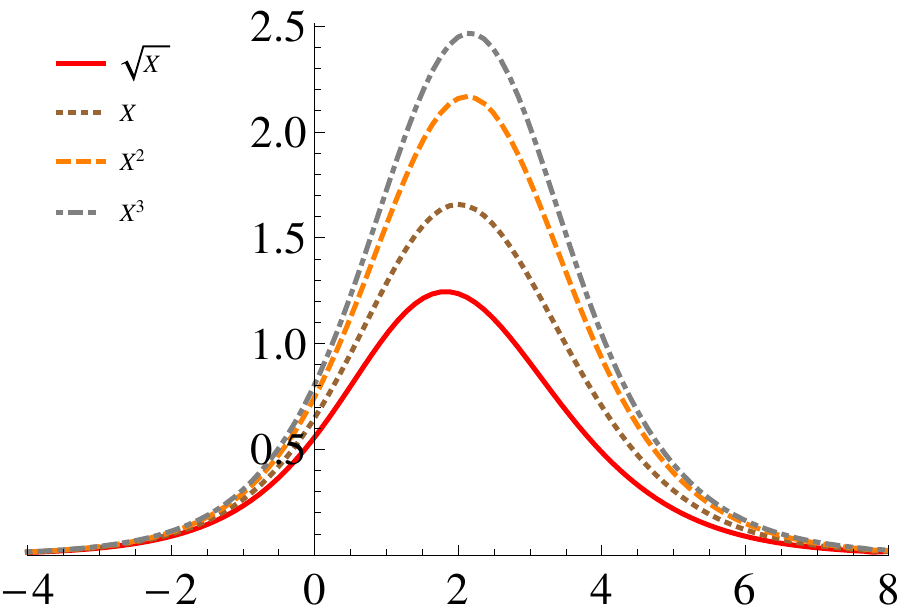}
 \end{subfigure} \qquad \qquad \qquad
 \begin{subfigure}
  \centering
  \includegraphics[width=0.3\textwidth]{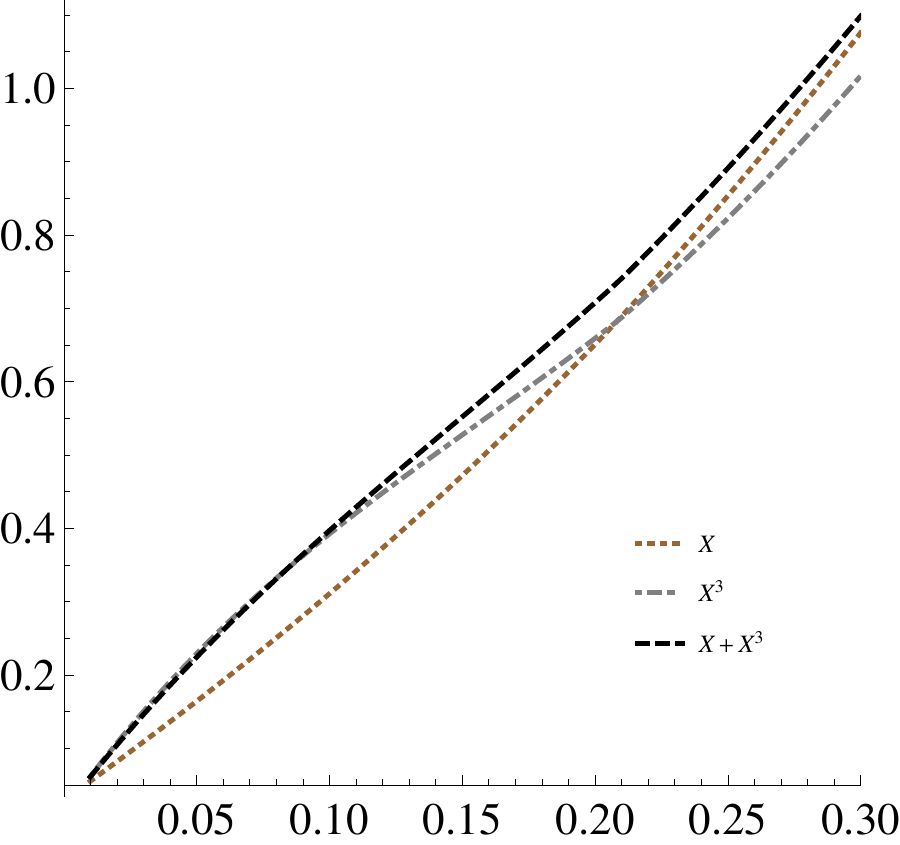}
 \end{subfigure} \\ \vskip+1em
 \centering
 \begin{subfigure}
  \centering
  \includegraphics[width=0.4\textwidth]{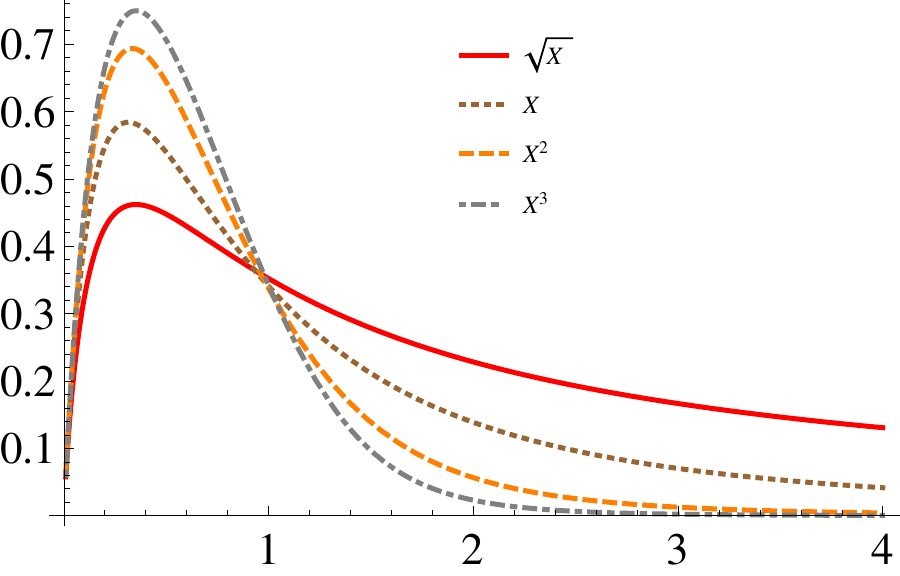}
 \end{subfigure} \qquad \qquad
 \begin{subfigure}
  \centering
  \includegraphics[width=0.4\textwidth]{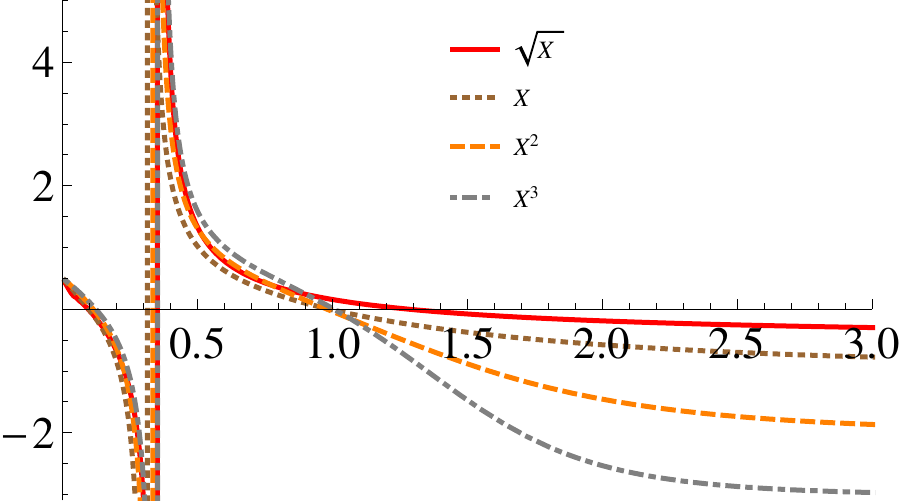}
 \end{subfigure}
 \begin{picture}(100,0)
  \put(19,66){\small{$\delta \gamma$}}
  \put(48,40){\small{$\ln \left(\frac{E_{\mathrm{static}}}{T_{(0)}} \right)$}}
  \put(65,67){\small{$\gamma$}}
  \put(95,39){\small{$\beta$}}
  \put(6,33){\small{$\delta \gamma$}}
  \put(45,7){\small{$T$}}
  \put(54,30){\small{$N_{\mathrm{scaling}}=\frac{1}{2}\left(1+T \frac{\delta \gamma''(T)}{\delta \gamma'(T)}\right)$}}
  \put(97,13){\small{$T$}}
 \end{picture}
 \vskip-2em
 \caption{Various plots related to the drag coefficients for the monomial and polynomial potentials. The potentials displayed are vanishing ($V(X)=0$, dashed blue), the square root ($V(X)=\sqrt{X}$, solid red), linear ($V(X)=X$, dotted brown), quadratic ($V(X)=X^2$, dashed orange), cubic ($V(X)=X^3$, dot-dashed grey) and mixed ($V(X)=X+X^3$, solid black) \textbf{Top left:} The drag coefficient minus its zero graviton mass value as a function of $E_{\mathrm{static}}$ for various potentials with $T=1$ and $\beta=1,2,3$. Note that in both extreme limits of $E_{\mathrm{static}}$ the difference goes to zero in accordance with \eqref{Eq:Drag}. The drag coefficient minus its zero graviton mass value as a function of temperature for various monomial potentials with $E_{\mathrm{static}}=1$ and $\beta=1$. \textbf{Top right:} The drag coefficient for the non-linear potential $V(X)=X+ \kappa X^3$ with $\kappa/m^4 =1$, $E_{\mathrm{static}}=1$ and $\beta=1$. We note that for low temperature the coefficient asymptotes the 
highest power behaviour of the potential; while for large temperature its behaviour is governed by the lowest power of the potential. This feature is quite universal and it is the inverse of the $\beta$ dependence case where the highest power dominates in the large $\beta$ limit. \textbf{Bottom Left:} The difference in the drag coefficient from it's zero gravitom mass value against temperature for various potentials with $E_{\mathrm{static}}=1$ and $\beta=5$. \textbf{Bottom Right:} A plot demonstrating the scaling behaviour of the drag coefficient in temperature for various monomial potentials with $E_{\mathrm{static}}=1$ and $\beta=5$. For large temperature the drag coefficient has a behaviour $\sim T^{-2N}$ which is reflected in the fact that the $N_{\mathrm{scaling}} \sim N$ at large temperatures.}
\label{fig:DampingAllPotentials}
\end{figure}
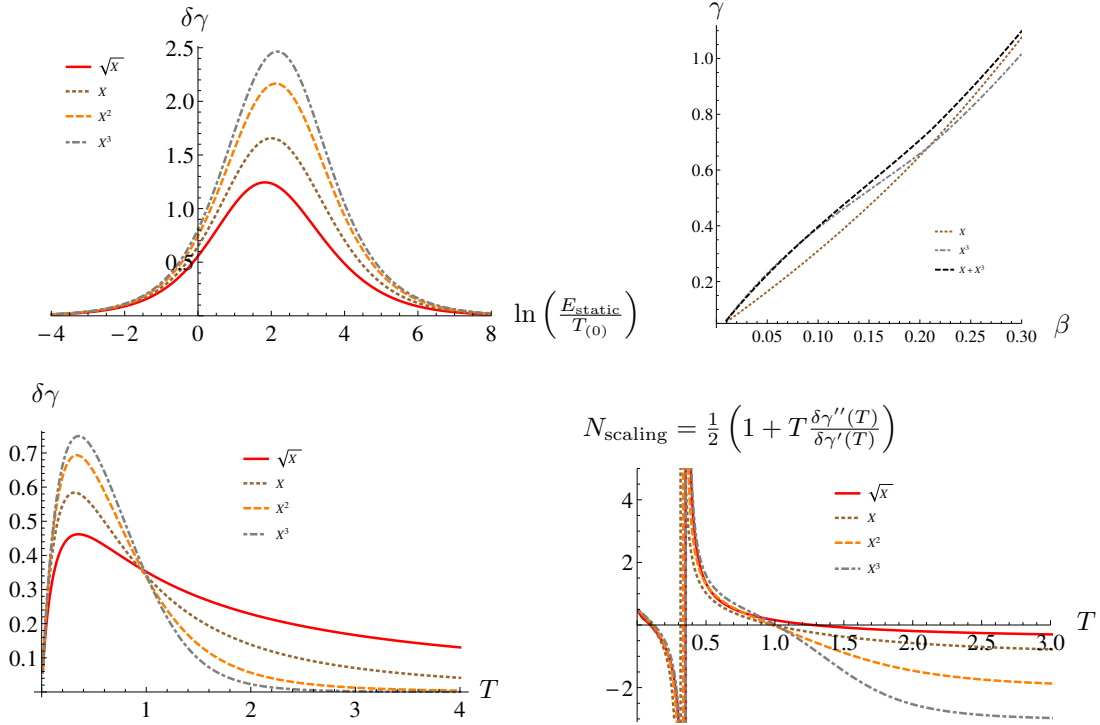

{\ We display the results of numerical computations of the drag coefficient for various potentials in fig.~\ref{fig:DragCoefficients} and fig.~\ref{fig:DampingAllPotentials}. We begin first with vanishing graviton mass summarising the results of \cite{Gubser:2006bz,Herzog:2006gh}. For small static energies the decay constant of the string is linear in the temperature and independent of $E_{\mathrm{static}}$,
  \begin{eqnarray}
   \label{Eq:DragZeroMassSmallEstatic}
   \left. \gamma \right|_{m=0} = 2 \pi T \; . 
  \end{eqnarray}
For large static energies the decay constant is 
  \begin{eqnarray}
   \label{Eq:DragZeroMassLargeEstatic}
   \left. \gamma \right|_{m=0} = \frac{\frac{4 \pi T}{3}}{1 + \frac{3}{4 \pi T} \left( \frac{E_{\mathrm{static}}}{T_{(0)} } \right)} \sim \frac{T^2}{E_{\mathrm{static}}} \; . 
  \end{eqnarray}
On fixing the temperature and varying the static energy we see in the top left plot of fig.~\ref{fig:DragCoefficients} a smooth interpolation between a constant and $1/E_{\mathrm{static}}$ behaviour at small and large $E_{\mathrm{static}}$ respectively.}

{\ Turning now to the linear potential, in the bottom plot of fig.~\ref{fig:DragCoefficients} we display the drag coefficient for various values of the graviton mass, with $E_{\mathrm{static}}/T_{(0)} =1$, against temperature normalised by the zero graviton mass result. The difference is always positive indicating that there is an increased drag when the graviton mass is non-zero. As expected when the temperature increases this effect diminishes as the spacetime begins to resemble thermal AdS. When the temperature is small with respect to both $\beta$ and $E_{\mathrm{static}}$ we find that the drag coefficient tends to a constant. It then increases with increasing temperature before peaking and subsequently decaying in a manner consistent with the high temperature result. In the top right plot of fig.~\ref{fig:DragCoefficients} we can see that as $\beta$ increases the drag coefficient also increases. The amount that the drag coefficient grows by is dependent upon the static energy of the string with smaller 
energies and larger energies being less affected by the graviton mass.}

{\ In the top right plot of fig.~\ref{fig:DampingAllPotentials} we display the drag coefficient for various potentials against $\beta$. Again we see that as $\beta$ increases the drag coefficient also increases. When $\beta$ vanishes we get the zero graviton mass value for the drag coefficient which is generally dependent on the static energy and the temperature. For small values of $\beta$ the larger the power of the monomial potential the larger the drag coefficient. This switches at some finite value of $\beta$ so that potentials with greater powers have smaller drag coefficients.}

{\  From the top left plot of fig.~\ref{fig:DampingAllPotentials} we also note that at large or small $E_{\mathrm{static}}$ the dependence of the drag coefficient on the potential diminishes just as was seen in the top right plot of fig.~\ref{fig:DragCoefficients}. In the bottom left plot of fig.~\ref{fig:DampingAllPotentials} we another generic feature of the drag coefficient - a peak in the temperature. It is interesting that the existence of this peak is a generic fingerprint of the model which does not rely strongly upon the choice of a particular potential. Additionally, for sufficiently large temperature, the difference in the drag coefficient scales as $T^{-2N}$ (where $N$ is the power of the monomial potential) as shown in bottom-right of ~\ref{fig:DampingAllPotentials}.}

{\ Having discussed our numerical results we would like to interpret them. For small static energies at non-zero temperature we use the analytic results of appendix \ref{sec:decayanalytic}, namely \eqref{Eq:Shortstringgamma}, to show that graviton mass has no effect on the decay constant. This is consistent with the top right and top left plots in fig.~\ref{fig:DragCoefficients} and with $\delta \gamma$ going to zero for small $E_{\mathrm{static}}$ in the top left plot of fig.~\ref{fig:DampingAllPotentials}. In terms of the gravity theory this is unsurprising because the near horizon geometry of \eqref{Eq:LineElement} takes the same form for both thermal AdS and all our massive gravity spacetimes at non-zero temperature. The interpretation in the field theory is also relatively straightforward - namely the static energy is much less than the average thermal energy and thus the string is excited to the average thermal energy independent of the small differences in mass.}

{\ For small and large static energies we use \eqref{Eq:Longstringqnm} to find
  \begin{eqnarray}
    \label{Eq:Drag}
    \gamma &=& \left\{ \begin{array}{ccc}
                        2 \pi T \; , & & E_{\mathrm{static}} \ll T \\
			\frac{1}{z_{\mathrm{H}}^2(\beta,T) E_{\mathrm{static}}} \; , & & E_{\mathrm{static}} \gg T
                       \end{array} \right. \; . 
  \end{eqnarray}
Clearly in the limit of large temperatures compared to $\beta$, where the graviton mass becomes negligible, this reproduces \eqref{Eq:DragZeroMassSmallEstatic} and \eqref{Eq:DragZeroMassLargeEstatic}. More generally, for $E_{\mathrm{static}} \gg T$, $\gamma \sim s/ E_{\mathrm{static}} \sim 1/(\mu(T) E_{\mathrm{static}})$ where the behaviour of the mobility is displayed for various potentials in fig.~\ref{fig:Mobility}. As we noted the mobility is bounded above at fixed temperature by the zero graviton mass result and is a decaying function as $\beta$ is increased and temperature decreased. Similarly therefore the decay constant of strings with large static energies is bounded below at fixed temperature and generically increases with decreasing $\beta$ and increasing temperature.}

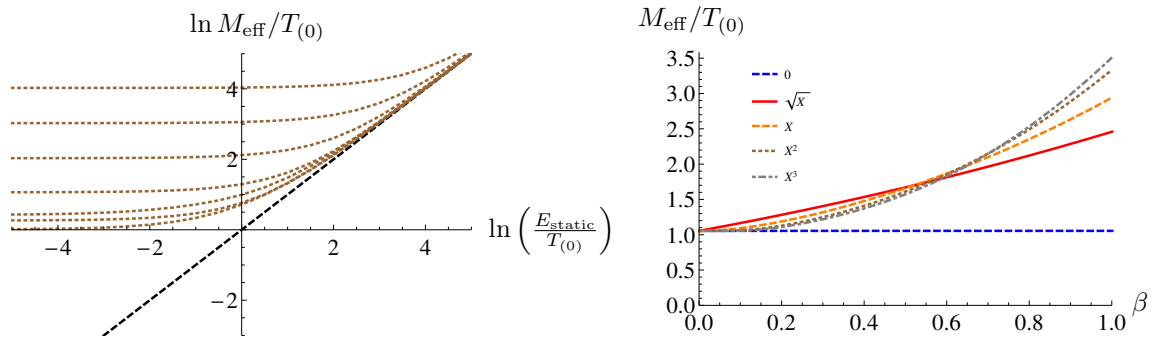
\begin{figure}[t]
 \centering
 \begin{subfigure}
  \centering
  \includegraphics[width=0.4\textwidth]{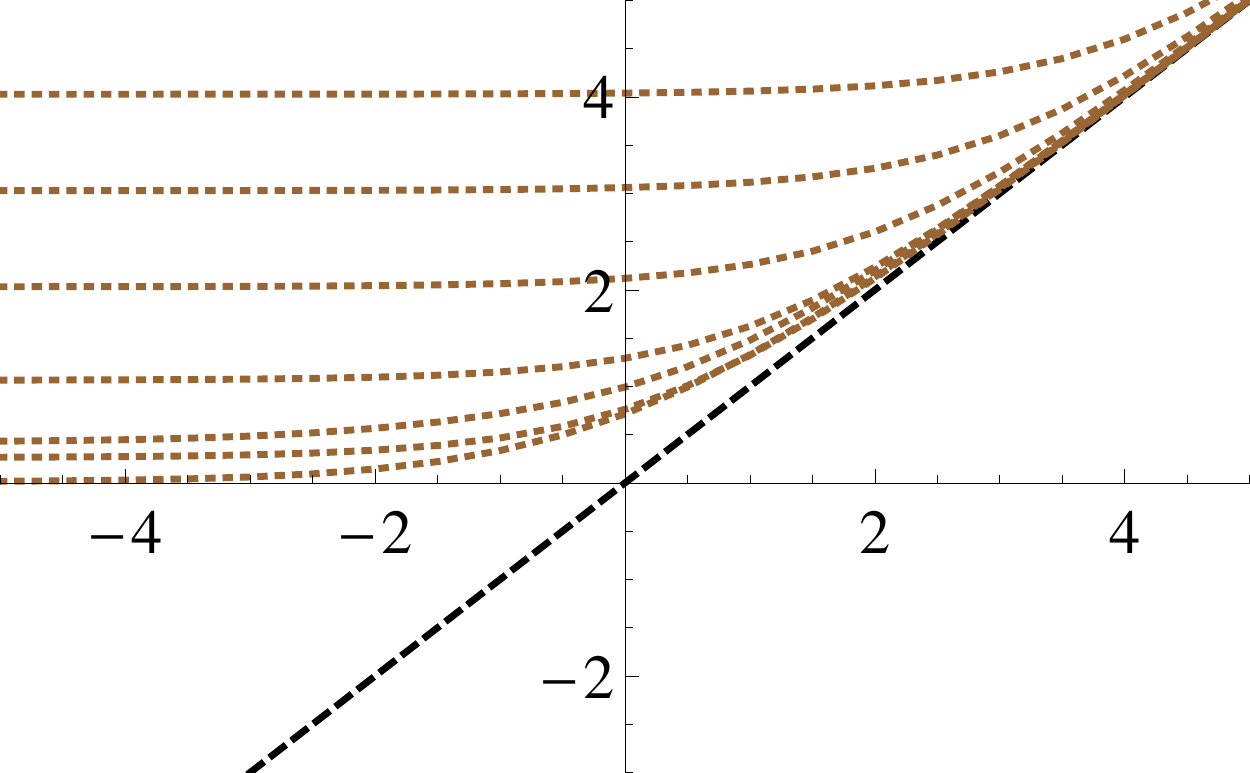}
 \end{subfigure} \qquad \qquad \qquad
 \begin{subfigure}
  \centering
  \includegraphics[width=0.4\textwidth]{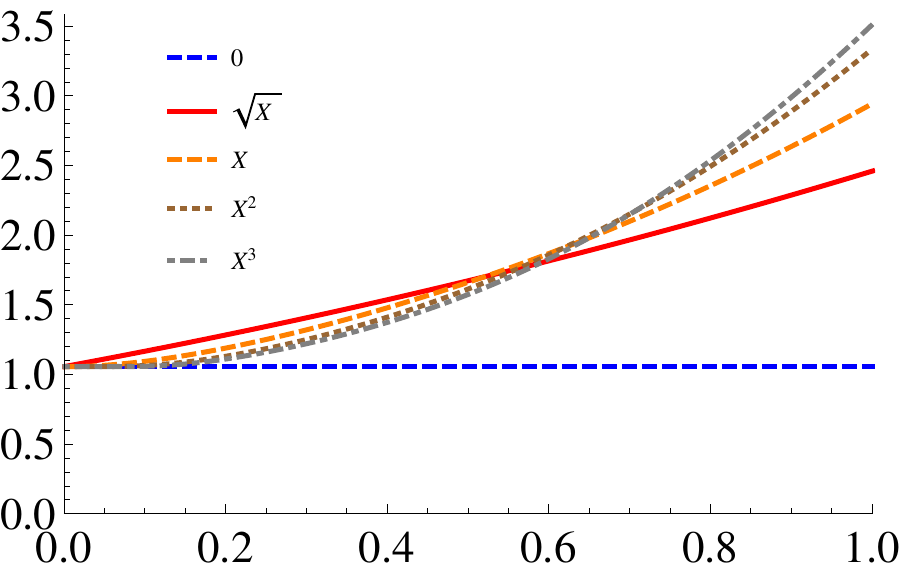}
 \end{subfigure} 
 \begin{picture}(100,0)
  \put(17,31.5){\small{$\ln M_{\mathrm{eff}}/T_{(0)}$}}
  \put(43,13.5){\small{$\ln \left(\frac{E_{\mathrm{static}}}{T_{(0)}}\right)$}}
  \put(56,32){\small{$M_{\mathrm{eff}}/T_{(0)}$}}
  \put(99,7){\small{$\beta$}}
 \end{picture}
 \vskip-2em
 \caption{Various plots for the effective kinetic mass. The potentials displayed are vanishing ($V(X)=0$, straight dashed blue line), square root ($V(X)=\sqrt{X}$, solid red), linear ($V(X)=X$, dotted brown), quadratic ($V(X)=X^2$, dashed orange) and cubic ($V(X)=X^3$, dot-dashed grey) \textbf{Left:} The logarithm of the effective kinetic mass of the string against the logarithm of the static energy in the background of linear potential, $V(X)=X$, for $\beta=1$ and $\ln T= -3,-2,\ldots,2,3$ (bottom brown dotted line to top). At large and small $E_{\mathrm{static}}$ with respect to temperature we see the result \eqref{Eq:AnalyticEffectiveMass}. \textbf{Right:} The effective mass a string against $\beta$ for various potentials with $T=0.04$.}
\label{fig:EffectiveMass}
\end{figure}

{\ For completeness we also determine the effective kinetic mass. In the limits of small and large static energy it has the analytic form:
  \begin{eqnarray}
       \label{Eq:AnalyticEffectiveMass}
       \frac{ M_{\mathrm{eff}} }{T_{(0)} }
   &=& \left\{ \begin{array}{ccc} \frac{1}{2 \pi T z_{\mathrm{H}}^2(\beta,T)}, & & E_{\mathrm{static}} \ll T  \\
				 E_{\mathrm{static}} , & & E_{\mathrm{static}} \gg T \end{array} \right. \; .
  \end{eqnarray}
This is compared to numerics in fig.~\ref{fig:EffectiveMass}. The left-hand plot shows the dependence of the effective kinetic mass upon the static energy for the linear potential $V(X)=X$ at various temperatures and fixed $\beta=1$. It is clear that as the static energy is increased we get the linear result given in \eqref{Eq:AnalyticEffectiveMass}. As the temperature is increased in the limit of small static energies the effect of $\beta$ is diminished and we get the analytic result
  \begin{eqnarray}
       \frac{ M_{\mathrm{eff}} }{T_{(0)} }
   &=& \frac{8 \pi}{9} T \; . 
  \end{eqnarray}
This can be seen in the left hand plot of fig.~\ref{fig:EffectiveMass} by the fact that for large temperature the brown lines at small $E_{\mathrm{static}}$ have approximately linear spacing. The right hand plot of fig.~\ref{fig:EffectiveMass} shows the effect of varying $\beta$ for the monomial potentials at fixed $T=0.04$ and $E_{\mathrm{static}}=1$. Above some value of $\beta$ the larger the power in the monomial, the larger the effective mass. Because we have chosen a relatively large $E_{\mathrm{static}}$ compared to the temperature as $\beta$ is taken to zero the monomial potentials have a kinetic mass which is almost equal to the kinetic mass of the string at vanishing graviton mass.}

\section{Conclusions}

\begin{table}[!h]
  \tiny
  \renewcommand{\arraystretch}{2}%
  \centering
  \begin{tabular}{|c|ccc|c|} \hline
		  & \multicolumn{3}{|c|}{Broken translation invariance}	 & Unbroken translation invariance \\ \cline{2-4}
		  & \multicolumn{1}{|c|}{$V(X) = \sqrt{X} $} & \multicolumn{1}{|c|}{$V(X) = X$} & $V(X)=X^{n}, \; n>2$ & $V(X)=0$ \\ \hline
   $c_{v}/\left. c_{v} \right|_{m=0}$ 
	   & \multicolumn{1}{|c|}{$ \left( 1 + \frac{\beta}{4 \pi T} \right) $}
	   & \multicolumn{1}{|c|}{$ \left( \frac{1 + 6 \left(\frac{\beta}{4 \pi T}\right)^2 + \sqrt{1+12 \left(\frac{\beta}{4 \pi T}\right)^2}}{2 \sqrt{1+ 12 \left(\frac{\beta}{4 \pi T}\right)^2}} \right)$}
	   & $\stackrel{T \rightarrow \infty}{\longrightarrow} 1 $
	   & $\left. c_{v} \right|_{m=0} = \frac{2 (4 \pi)^3 T^2}{3^2} M_{P}^2$ \\ \hline
   $\mu/\mu_{\mathrm{bound}}(T)$   
	   & \multicolumn{1}{|c|}{$  \left( 1 + \frac{\beta}{4 \pi T} \right)^{-2}$}  
	   & \multicolumn{1}{|c|}{$ \left( \frac{\sqrt{ 1 + 12 (\frac{\beta}{4 \pi T})^2 } - 1}{6 \left( \frac{\beta}{4 \pi T} \right)^2} \right)^2$} 
	   & $ \sim \frac{1}{T^{2n}} $
	   & $ \mu_{\mathrm{bound}}(T) =\frac{1}{T_{(0)} } \left( \frac{3}{4 \pi T} \right)^2 $ \\ \hline
   $\gamma$ 
	   & \multicolumn{1}{|c|}{\multirow{2}{*}{-}} 
	   & \multicolumn{1}{|c|}{\multirow{2}{*}{-}}
	   & \multirow{2}{*}{$2 \pi T$}
	   & \multirow{2}{*}{$\left. \gamma \right|_{m=0} = 2 \pi T$} \\
   $(E_{\mathrm{static}} \ll T)$ & \multicolumn{1}{|c|}{} & \multicolumn{1}{|c|}{} & \multicolumn{1}{|c|}{} & \\ \hline
   $\gamma$ 
	   & \multicolumn{1}{|c|}{\multirow{2}{*}{-}} 
	   & \multicolumn{1}{|c|}{\multirow{2}{*}{-}}
	   & \multirow{2}{*}{$\sim \frac{T^{2n}}{E_{\mathrm{static}}}$}
	   & \multirow{2}{*}{$\left. \gamma \right|_{m=0} = \frac{\frac{4 \pi T}{3}}{1 + \frac{3}{4 \pi T} \left( \frac{E_{\mathrm{static}}}{T_{(0)} } \right)}$} \\
   $(E_{\mathrm{static}} \gg T)$ & \multicolumn{1}{|c|}{} & \multicolumn{1}{|c|}{} & \multicolumn{1}{|c|}{} & \\ 
   \hline
  \end{tabular}
  \caption{A representative sample of the results available in this paper. $c_{v}$ is the volumetric heat capacity, $\mu$ is the mobility, $\gamma$ the drag coefficient, $T$ the temperature, $T_{(0)}$ the string tension, $E_{\mathrm{static}}$ the static energy of the string, $m$ the graviton mass, $M_{P}$ the Planck mass and $\beta$ a model dependent parameter which can be chosen freely. Results for more generic potentials can be found in the paper.}
  \label{Tab:summary}
\end{table}

{\ We have calculated the momentum loss rates of probe strings in various solutions to a particular, very general, model of massive gravity. We have found that the mobility and subsequently the diffusion constant of the dual point particles is bounded above by the zero graviton mass result. Moreover we have extracted the contribution of the graviton mass to the decay rate of a particle given some initial impulse. These results are summarised in table \ref{Tab:summary}.}

{\ We have argued that at low temperatures the behaviour of the string is governed by the existence of a set of non-trivial ground states. This is of course predicated on the idea that these ground states, which have non-zero entropy density at zero temperature, are stable. However, it is known that often such states are unstable. As these new ground states are quite exotic they are worthy of independent study which would simultaneously determine whether we can trust our results at low temperatures.}

{\ We have seen that it is not possible to assign the additional momentum loss felt by a probe particle in field theories dual to massive gravity to a unique condensed matter phenomenon such as phonons or impurities. Our work thus indicates that extreme caution must be employed when using condensed matter terminology to describe the outcome of holographic experiments with massive gravity spacetimes. However, it has been argued \cite{Blake:2013owa} that a small graviton mass can be generated by a weak lattice in a gravitational version of the Higgs' mechanism. Understanding the relationship between a probe string moving in a background with a non-weak lattice and the subsequent effective massive gravity theory should in principle  help to clarify the role of different mechanisms, such as phonon production and interactions with impurities, responsible for the additional momentum loss observed in this paper.}

{\ Finally, we made the simplifying assumption that our classical string was uncharged under the scalar fields. If the total action given by the sum of \eqref{Eq:MassiveGravityAction} and the string action is to preserve the shift symmetry $\Phi^{I} \rightarrow \Phi^{I} + \alpha^{I}$, so that we can treat the $\Phi^{I}$-fields as Goldstone bosons for broken translation invariance, the action of the string can have the more general form
  \begin{eqnarray}
    \label{Eq:StringAction2}
    S = - T_{(0)} \int d^{2} \sigma \; e^{f^{(1)}(\partial \Phi)} \sqrt{ - \mathrm{det} \left( g_{MN}(X) \partial_{a} X^{M} \partial_{b} X^{N} + f^{(2)}_{ab}(\partial \Phi) \right) } \; , 
  \end{eqnarray}
where $f^{(1)}(\partial \Phi)$ and $f^{(2)}_{ab}(\partial \Phi)$ are two unknown functions of $\partial_{a} \Phi^{I}$. The function $f^{(2)}_{ab}(\partial \Phi)$ describes how the string is charged under the fields $\Phi^{I}$ while $T_{(0)} \exp(f^{(1)}(\partial \Phi))$ describes some effective tension. It is also possible to add a Fradkin-Tseytlin term to the action\footnote{We thank the anonymous reviewers for pointing out that such a term could have non-trivial effects.}. Such a term would couple any non-trivial background dilaton to the worldsheet curvature and significantly complicate the equations of motion. As we have assumed that the background dilaton vanishes we have not worried about such a term. Even in the presence of a non trivial dilaton profile however two comments are in order: firstly, the Fradkin-Tseytlin term would probably be an $\alpha'$ correction, which is certainly interesting but beyond the scope of the present paper. Secondly it is not clear from the holographic point of view whether such a term contributes to the on-shell action or whether it represents a pure counterterm\footnote{We would like to thank E.~Kiritsis for discussions on this point.}. That said, it would certainly be interesting to understand the effects of such a term. }

{\ It may be interesting to ask how the physics changes when dependence on the scalar fields enters through more than the pullback of the bulk metric. We still expect to see non-zero momentum loss at zero temperature, if only due to the interaction of the string with the background scalars, although it is unclear to us whether our bound on the diffusion constant should survive. Moreover, it may be possible to motivate from a string model an effective field theory where the string is charged under the translation breaking scalars using methods similar to \cite{Blake:2013owa}. This would allow a precise interpretation of the string in terms of the boundary field theory as well as providing explicit forms for the undetermined functions in our string action.}

\acknowledgments

{\ MB acknowledges support from MINECO under grant FPA2011-25948, DURSI under grant 2014SGR1450 and Centro de Excelencia Severo Ochoa program, grant SEV-2012-
0234. DB is supported by the Israeli Science Foundation (ISF) 392/09 and a Fine Fellowship. We would like to thank Oren Bergman, Mike Blake, Alessandro Braggio, Mikhail Goykhman,  E.~Kiritsis, Oriol Pujol\'as and Amos Yarom for reading early drafts and offering various insights into the problems presented here. MB would additionally like to thank Federico Caglieris,  Giacomo Dolcetto, Francesca Telesio and Genis Torrents for useful conversations and comments.}

\appendix

\section{Analytic results for the mobility}
\label{sec:mobilityanalytic}

{\ Consider the fluctuation equation \eqref{Eq:Fluctuationeqn}. Shifting to Fourier space with the conventions, 
  \begin{eqnarray}
   \label{Eq:Fourierfluc}
   \delta x^{1}(z,t) = \int \frac{d \omega}{2 \pi} \; \delta x^{1}(z;\omega) \exp( i \omega t ) \; , 
  \end{eqnarray}
we find the spatial momentum of the slow string is
  \begin{eqnarray}
   p(\omega) = \int_{z=z_{\mathrm{max}}}^{z_{\mathrm{m}}} dz \; \pi^{t}_{x}(z;\omega) = \epsilon \left[ \frac{T_{(0)} }{i \omega}  \frac{f(z_{\mathrm{max}})}{z_{\mathrm{max}}^2} \partial_{z} \delta x^{1}(z_{\mathrm{max}};\omega) \right] + \mathcal{O}^{3}(\epsilon)  \; , 
  \end{eqnarray}
where we have regulated the momentum in the IR with an explicit cut-off $z_{\mathrm{max}}$ which in principle we should take to be $z_{\mathrm{H}}$. Satisfying the outgoing condition at the past horizon requires $f(z_{\mathrm{max}}) \partial_{z} \delta x^{1}(z_{\mathrm{max}};\omega) = i \omega \delta x^{1}(z_{\mathrm{max}};\omega)$ as $z_{\mathrm{max}} \rightarrow z_{\mathrm{H}}$. Integrating $p(\omega)$ against frequency we see
  \begin{eqnarray}
    p(t) = \int \frac{d\omega}{2 \pi} \; p(\omega) \exp( i \omega t )  
      = \frac{T_{(0)} }{z_{\mathrm{max}}^2} \delta x^{1}(z_{\mathrm{max}},t) \; .
  \end{eqnarray}
}

{\ Turning now to the energy density of the string,
  \begin{eqnarray}
    \pi^{t}_{t}(z,t) = - \frac{ T_{(0)}  }{ z^2 } \left[ 1 + \epsilon^2 \left( \frac{(\partial_{t} \delta x^{1}(z,t))^2}{2 f(z)} + \frac{f(z)}{2} (\partial_{z} \delta x^{1}(z,t))^2 \right) + \mathcal{O}^{4}(\epsilon) \right]\; ,
  \end{eqnarray}
and integrating against the string length we find the total energy of the string is equal to 
  \begin{eqnarray}
   \label{Eq:PreDispersionRelation}
   E = E_{\mathrm{static}}(T) + \epsilon^2 \frac{T_{(0)} }{2 z_{\mathrm{max}}^2} \left[ \delta x^{1}(z_{\mathrm{max}},t) \partial_{t} \delta x^{1}(z_{\mathrm{max}},t) \right] + \mathcal{O}^{4}(\epsilon)
  \end{eqnarray}
where we have imposed the outgoing condition $f(z_{\mathrm{max}}) \partial_{z}  \delta x^{1}= \partial_{t}  \delta x^{1}$ as $z_{\mathrm{max}} \rightarrow z_{\mathrm{H}}$.}

{\ If the free string endpoint is to satisfy the drag equation \eqref{Eq:DragEquation} it must be the case that $p(t)=p(0) \exp(-\gamma t)$. Using the low velocity limit and integrating the velocity against time implies $\delta x^{1}(t)=\delta x^{1}(0) \exp(-\gamma t)$ up to an additive constant we set to zero. Substituting into \eqref{Eq:PreDispersionRelation} then gives
  \begin{eqnarray}
    \label{Eq:DispersionRelationAppendix}
    E = E_{\mathrm{static}}(T) + \frac{\vec{p}^2}{2 M_{\mathrm{eff}}(T)} + \mathcal{O}^{4}(\vec{p}) \; , \qquad \mu^{-1} = \gamma M_{\mathrm{eff}} = \frac{T_{(0)} }{z_{\mathrm{H}}^2} \; ,
  \end{eqnarray}
with $E_{\mathrm{static}}$ given by \eqref{Eq:Staticenergy}.We have yet to show that we can arrange for the endpoint to move according to $p(t)=p(0) \exp(-\gamma t)$, and determine the drag coefficient $\gamma$, which we shall do this in section \ref{Sec:QNM}.}

\section{Analytic results for the decay constant}
\label{sec:decayanalytic}

{\ We can extract some analytical behaviour for our drag coefficients in the regimes of small and large static energies. Beginning with the former we solve the fluctuation equation \eqref{Eq:QNMequation} in a near horizon expansion and find the derivative of the fluctuation is
  \begin{eqnarray}
   \partial_{z} \delta x^{1} = - \frac{1}{z_{\mathrm{H}}} 
			       \left( \frac{z_{\mathrm{H}} - z_{\mathrm{m}}}{z_{\mathrm{H}}} \right)^{- \frac{i \omega}{4 \pi T} -1} 
			       \left[ \pm \frac{i \omega}{4 \pi T} 
				      + \left(1 - \frac{i \omega}{4 \pi T} \right) c_{1} (z_{\mathrm{H}}-z_{\mathrm{m}}) 
				      + \mathcal{O}^{2}(z_{\mathrm{H}}-z_{\mathrm{m}}) \right] \nonumber
  \end{eqnarray}
where
  \begin{eqnarray}
   c_{1} &=& i \omega \left[ \frac{32 (\pi T)^2 + z_{\mathrm{H}} f''(z_{\mathrm{H}}) \left(2 \pi T - i \omega \right)}
				      {32 (\pi T)^2 z_{\mathrm{H}} \left( 2 \pi T - i \omega \right)} \right] \; . 
  \end{eqnarray}
Setting $z_{\mathrm{m}}=z_{\mathrm{H}}$ imposes a constraint on the frequency and we find
  \begin{eqnarray}
   \label{Eq:Shortstringgamma}
   \omega = 2 \pi T i \; . 
  \end{eqnarray}
This is the same result one finds for a short string in thermal AdS \cite{Gubser:2006bz,Herzog:2006gh}.}

{\ A second regime where we can find an analytic solution to compare with the numerics is the long string. To solve the fluctuation equation \eqref{Eq:QNMequation} assume that we can perform a small frequency expansion for non-trivial $\omega$,
  \begin{eqnarray}
   \label{Eq:QNMomegaexpansion}
   \delta x^{1}(z;\omega) &=& \delta x^{(0)}_{1}(z;\omega) + \omega \delta x^{(1)}_{1}(z;\omega) + \omega^2 \delta x^{(2)}_{1}(z;\omega)
			      + \mathcal{O}^{3}(\omega) \; .
  \end{eqnarray}
We will find that non-trivial solutions exist only when $\omega$ is of order $z_{\mathrm{m}}$ with $z_{\mathrm{m}} \ll 1$. We impose the outgoing condition on the past horizon and use an overall scaling of the linerised fluctuation equations
  \begin{eqnarray}
   \label{Eq:QNMequationlong}
   0 &=& \left[ z^2 f(z) \partial_{z} \left( \frac{f(z)}{z^2} \partial_{z} \delta x^{1}_{(0)}(z;\omega) \right) \right]
	 + \omega \left[ z^2 f(z) \partial_{z} \left( \frac{f(z)}{z^2} \partial_{z} \delta x^{1}_{(1)}(z;\omega) \right) \right] \nonumber \\
     &\;& + \omega^2 \left[ z^2 f(z) \partial_{z} \left( \frac{f(z)}{z^2} \partial_{z} \delta x^{1}_{(2)}(z;\omega) \right) + \delta x^{1}_{(0)}(z;\omega) \right] 
	  + \mathcal{O}^{3}(\omega) \; ,
  \end{eqnarray}
to set 
  \begin{eqnarray}
   \lim_{z \rightarrow z_{\mathrm{H}}} \left[ \left(\frac{z_{\mathrm{H}}-z}{z_{\mathrm{H}}}\right)^{ \frac{i \omega}{4 \pi T}} \delta x^{1}(z;\omega) \right] = 1 \; . 
  \end{eqnarray}
Solving the equations \eqref{Eq:QNMequationlong} up to second order in $\omega$ and calculating the derivative at $z_{\mathrm{m}}$ we find (logarithmic term is not quite right)
  \begin{eqnarray}
      \left. \partial_{z} \delta x^{1}(z;\omega) \right|_{z=z_{\mathrm{m}}} 
   &=& \omega \left[ - \frac{i}{z_{\mathrm{H}}^2} \frac{z_{\mathrm{m}}^2}{f(z_{\mathrm{m}})} \right] \nonumber \\
   &\;& + \omega^2 \left[ \frac{z_{\mathrm{m}}^2}{f(z_{\mathrm{m}})} \lim_{z \rightarrow z_{\mathrm{H}}} 
	\left( \int^{z}_{w=z_{\mathrm{m}}} dw \; \frac{1}{w^2 f(w)}  + \frac{1}{4 \pi T z_{\mathrm{H}}^2} 
	\ln\left(\frac{z_{\mathrm{H}}-z}{z_{\mathrm{H}}}\right) \right) \right] \nonumber \\
   &\;& + \mathcal{O}^{3}(\omega) \; . 
  \end{eqnarray}
It is clear that for a non-trivial solution we cannot truncate to order $\omega$ but must cancel the order $\omega$ and $\omega^2$ terms. Expanding the Neumann condition to leading order in small $z_{\mathrm{H}}-z_{\mathrm{m}}$ we see find 
  \begin{eqnarray}
    \label{Eq:Longstringqnm}
    \omega = i z_{\mathrm{m}}/z_{\mathrm{H}}^2 \; . 
  \end{eqnarray}
This is small, as required by our expansion \eqref{Eq:QNMomegaexpansion}, on the condition that the string is long i.e.~$z_{\mathrm{m}} \ll 1$.}

{\ We see that the long string has fixed dependence on $z_{\mathrm{m}}$ and $z_{\mathrm{H}}$ as long as the line element has the form \eqref{Eq:LineElement}. The dependence on $z_{\mathrm{m}}$ and $z_{\mathrm{H}}$ is the same as that found by \cite{Gubser:2006bz,Herzog:2006gh}. However, once again, because $z_{\mathrm{H}}$ depends on the temperature and graviton mass the drag coefficient of long strings is distinct between different theories of massive gravity at the same temperature.

\bibliography{Massive_gravity}
\bibliographystyle{JHEP}

\end{document}